\documentclass[a4paper,11pt]{article}
\pdfoutput=1 % if your are submitting a pdflatex (i.e. if you have
\usepackage{jheppub2} % for details on the use of the package, please
\usepackage[OT1]{fontenc} % if needed
\graphicspath{{figs/}} 	     %Set the default folder for images
\usepackage{booktabs} % To thicken table lines
\usepackage{bbold} % all ten digits with \mathbb
\usepackage{mathtools}
\usepackage{multirow}
\usepackage[version=4]{mhchem} % Chemical elements notation
\usepackage{physics}
\usepackage{pdflscape}
\usepackage{slashed}
\usepackage{xcolor}

\usepackage{tikz} % To draw things
\usepackage{tikz-feynman} % To draw Feynman-diagrams
\usepackage{tikz-feynhand} % To draw Feynman-diagrams
\tikzfeynmanset{compat=1.1.0} % To ensure compatibility / avoid unexpected errors

\usepackage{feynmp-auto}
\usepackage{graphicx}

\newcommand{\1}{{\bf 1}}
\newcommand{\2}{{\bf 2}}

\newcommand{\ol}[1]{\overline{#1}}
\newcommand{\wt}[1]{\widetilde{#1}}

\newcommand{\M}{{\cal M}}

\renewcommand{\d}{{\rm d}}
\def\arctanh{\mbox{arctanh}}

\newcommand{\SM}{{\rm SM}}
\newcommand{\hc}{{\rm h.c.}}

\newcommand{\GeV}{{\rm GeV}}

\newcommand{\MeV}{{\rm MeV}}

\def\beq#1\eeq{\begin{align}#1\end{align}}
\newcommand{\bea} {\begin{eqnarray}}
\newcommand{\eea} {\end{eqnarray}}

\renewcommand{\[}{\left[}
\renewcommand{\]}{\right]}
\renewcommand{\(}{\left(}
\renewcommand{\)}{\right)}

\begin{document}

\preprint{KYUSHU-HET-342}
% \preprint{YITP-25-YYY}
%\arxivnumber{2406.XXXXX}

%\def\ihalf{\nicefrac{i}{2}}\title{Electroweak symmetry (non)-restoration with improved thermal resummation}
% \def\ihalf{\nicefrac{i}{2}}
\title{Novel bounds on neutrino portal dark matter from leptonic meson decays}

\author[a,b,c]{Shohei Okawa,}
\author[d]{Yuji Omura}

\affiliation[a]{Asia Pacific Center for Theoretical Physics, Pohang, 37673, Korea}
\affiliation[b]{Department of Physics, Pohang University of Science and Technology, Pohang, 37673, Korea}
\affiliation[c]{Department of Physics, Kyushu University, 744 Motooka, Nishi-ku, Fukuoka, 819-0395, Japan}
% \affiliation[d]{Yukawa Institute for Theoretical Physics, Kyoto University, Kyoto 606-8502, Japan}
\affiliation[d]{Department of Physics, Kindai University, Higashi-Osaka, Osaka 577-8502, Japan}

\emailAdd{shohei.okawa@apctp.org}
\emailAdd{yomura@phys.kindai.ac.jp}

\abstract{
We investigate the potential of leptonic meson decays $M \to \ell \bar\nu_\ell$, 
where $M$ is a pseudo-scalar meson, as a probe of neutrino portal dark matter.
The model of our focus features a neutral fermion $\psi$ and scalar $\phi$, 
which are coupled predominantly to neutrinos in the form ${\cal L} \supset \lambda\,\ol\nu_L\,\phi\,\psi_R$. 
This interaction generates two corrections to the $M \to \ell \bar\nu_\ell$ observables. 
The first one is a novel three-body decay process $M \to \ell \bar\psi\phi$. 
This process is enabled by the splitting of the off-shell anti-neutrino $\bar\nu_\ell$ 
into $\psi$ and $\phi$ in the $M \to \ell \bar\nu_\ell$ diagram. 
The helicity suppression in $M \to \ell \bar\nu_\ell$ is absent in the three-body process, 
thereby forming a potentially large contribution to real experimental results, provided $\psi$ and $\phi$ are invisible.
The second one is one-loop radiative corrections to the weak vertex $W\ell \bar\nu_\ell$, 
which do not modify the charged lepton spectrum but lead to enhancement or suppression of the partial $M \to \ell \bar\nu_\ell$ decay width. 
To demonstrate the ability of the leptonic meson decays to probe the neutrino portal dark matter, 
we compute two corrections analytically and compare the modified meson branching ratios with 
the experimental data on the lepton flavor universality of pion and Kaon decays. 
The resulting constraints turn out to surpass the existing bounds in a large part of parameter spaces.
}

\maketitle

\renewcommand{\thefootnote}{\#\arabic{footnote}}
\setcounter{footnote}{0}
%%================================%%

\section{Introduction}

Despite the tremendous success of the Standard Model (SM), 
many theoretical mysteries and empirical deficiencies remain unsolved in particle physics. 
Dark matter (DM), in particular, represents a compelling signal for physics beyond the SM (BSM) and has been actively investigated. 
For a long time, the primary attention has been paid to neutral stable particles, 
with electroweak (EW) scale masses and weakly interacting to SM particles, 
also known as Weakly Interacting Massive Particles (WIMPs) \cite{Lee:1977ua, Hut:1977zn, Sato:1977ye, Vysotsky:1977pe, Bernstein:1985th}. 
The WIMP DM candidates not only are predicted in new physics models addressing the so-called hierarchy problem, 
but also can be naturally created from the SM plasma in the thermal history of the universe. 
Moreover, the very same interactions responsible for the thermal production can serve as a useful probe to 
directly test the WIMP DM scenario in terrestrial experiments. 
In the last decades, direct detection experiments have considerably improved the sensitivity to 
DM-nucleon scattering cross sections, albeit without any affirmative signals so far. 
These results provide very stringent constraints on a variety of WIMP DM models. 

Given the strong constraints on the traditional WIMP DM scenario, 
significant attention has recently been directed toward a light DM particle with a mass much below the EW scale. 
In this mass region, typically below the GeV scale, 
DM particles constituting the DM halo in the Milky Way do not have enough kinetic energy to 
leave observable signals at detectors, thereby evading the strong direct detection constraints. 
On the other hand, the thermal DM production in the early universe can still be achieved even in the sub-GeV mass region, 
accompanied with new light mediator particles \cite{Boehm:2003hm, Fayet:2004bw}. 

In most cases, realistic sub-GeV DM models comprise several new particles with masses much below the EW scale. 
Some particles are stable and serve as DM, while other particles act as force mediators to the SM sector. 
Among the popular mediators are dark photon \cite{Galison:1983pa, Holdom:1985ag, Holdom:1986eq, Foot:1991kb, Pospelov:2007mp, Arkani-Hamed:2008hhe}, dark Higgs \cite{Schabinger:2005ei, Patt:2006fw, Wells:2008xg, Batell:2009yf, Weihs:2011wp, Kim:2008pp, Pospelov:2011yp, Arkani-Hamed:2008hhe}, axion-like particles \cite{Nomura:2008ru, Dolan:2014ska, Gola:2021abm, Fitzpatrick:2023xks, Dror:2023fyd, Darme:2020sjf, Kamada:2017tsq, Hochberg:2018rjs, Bharucha:2022lty, Ghosh:2023tyz} and so on. 
These mediators realize the thermal DM production through an $s$-channel DM pair annihilation into electrons, muons, neutrinos, pions, and photons. 
The couplings of the mediators to the visible SM particles also offer a chance to (in)directly test this $s$-channel framework in medium-energy colliders and beam-dump experiments, 
as well as through modification of cosmological and astrophysical observables. 

Another viable realization of sub-GeV DM arises from introducing a light $t$-channel mediator \cite{Boehm:2013jpa, Batell:2017cmf, McKeen:2018pbb, Blennow:2019fhy, Biswas:2021kio, Li:2022bpp, Okawa:2020jea}. 
In this framework, DM pair annihilation into neutrinos is the leading process responsible for the thermal production. 
As the DM and mediator couple predominantly to neutrinos, 
experimental tests are more challenging than in the $s$-channel scenario. 
It has been pointed out that future large-scale neutrino telescopes could be utilized to indirectly probe this scenario, 
provided the DM annihilation proceeds in the $s$-wave \cite{Yuksel:2007ac, PalomaresRuiz:2007eu, Primulando:2017kxf, Campo:2017nwh, Campo:2018dfh, Klop:2018ltd, Arguelles:2019ouk, Bell:2020rkw, Asai:2020qlp, Akita:2022lit}.
However, this approach does not apply when the annihilation is velocity suppressed. 
Thus, complementary experimental probes are indispensable for a full exploration of such neutrino-philic DM candidates. 

In the present paper, we examine the potential of leptonic meson decays 
$M \to \ell\,\ol\nu_\ell$ ($M$ denotes a pseudo-scalar meson) as a probe of neutrino-philic DM. 
We focus on a class of DM models, where DM couples predominantly to neutrinos together with a light $t$-channel mediator. 
This model is analogous to the so-called neutrino portal DM models, characterized by the interaction ${\cal L}\supset \lambda\,\ol{\nu_L^{}}\,\phi\,\psi_R^{}$, where the lightest of $\phi$ and $\psi$ is DM and the other is a mediator. 
In this model, there are two corrections to the $M \to \ell\,\ol{\nu}_\ell$ observables. 
The first correction is a novel three-body decay process $M \to \ell\,\ol\psi\,\phi$. 
This process is enabled by the off-shell splitting of $\ol{\nu}_\ell$ into $\psi$ and $\phi$ in the $M \to \ell\,\ol{\nu}_\ell$ diagram, and will contribute to real experimental results as $\psi$ and $\phi$ are invisible.
The second one is one-loop radiative corrections to the weak vertex $W\ell\,\ol\nu_\ell$, which in turn modify the pertinent two-body decay process $M \to \ell\,\ol{\nu}_\ell$. 
We evaluate both corrections analytically and compare them with the latest experimental data. 

This paper is organized as follows. 
In Sec.~\ref{sec:model}, we present the neutrino portal DM interactions which are particularly focused on in this paper. 
The renormalizable extensions leading to the pertinent interactions at low energies are also discussed there.
In Sec.~\ref{sec:correction}, we explicitly analyze both three-body and one-loop corrections to the leptonic meson decays $M\to\ell\,\ol{\nu}_\ell$ in our model. 
In Sec.~\ref{sec:analysis}, we compare our calculation with the experimental results and derive the constraints. 
We briefly comment on other observables which one can potentially apply the same approach to in Sec.~\ref{sec:others}, 
and summarize our results and discuss the future direction in Sec.~\ref{sec:summary}.

\section{Setup}
\label{sec:model}

In this work, we consider a class of DM models, characterized by the so-called neutrino portal interactions, 
\begin{align}
{\cal L} & \supset - y_\ell\,\ol{\nu_L^{}}_{\ell}\,\phi\,\psi_R^{} +  {\rm h.c.},
\label{eq:Lint_eff}
\end{align}
where $\nu_{L\ell}^{}$ denote the SM left-handed neutrinos with $\ell=e,\mu,\tau$, and $\psi$ and $\phi$ are newly introduced neutral fermion and scalar, respectively. 
We assume a global $Z_2$ symmetry under which only $\psi$ and $\phi$ are odd while all the SM fields are even. 
The lighter one of $\psi$ and $\phi$ is thus a DM candidate and the other is a mediator. 
In this work, we consider $\psi$ is a vectorlike fermion having a Dirac mass term ${\cal L}\supset -m_\psi\bar{\psi}\psi$ and $\phi$ is a real scalar. 
This assumption is not necessary in general, and $\psi_R^{}$ can be Majorana and $\phi$ can be a complex scalar.

While eq.\,(\ref{eq:Lint_eff}) is not gauge invariant, it can effectively arise from renormalizable UV theories. 
To see it, we first recover the gauge invariance of this interaction term. 
If both $\psi$ and $\phi$ originate in gauge singlet fields, 
eq.\,(\ref{eq:Lint_eff}) can be extended to the dim-5 interaction, 
\beq
\ol{\nu_L^{}}_{\ell}\,\phi\,\psi_R^{} \to \frac{1}{\Lambda}\,\ol L_\ell\,\wt{\Phi}\,\phi\,\psi_R^{} \,,
\label{eq:dim5}
\eeq
where $L_\ell$ denotes the left-handed doublet leptons and $\wt{\Phi}=i\sigma_2 \Phi$ (with $\Phi$ the doublet Higgs field). 
This operator is still non-renormalizable but gauge invariant, and can reproduce eq.~(\ref{eq:Lint_eff}) after the electroweak (EW) symmetry breaking. 
Next, eq.\,(\ref{eq:dim5}) can be further deconstructed into two three-point interactions, accompanied by an additional heavy mediator: 
\begin{equation}
\frac{1}{\Lambda}\,\ol{L}_\ell\,\wt{\Phi}\,\phi\,\psi_R \to \left\{ 
\renewcommand{\arraystretch}{1.25}
\begin{array}{ll} 
\mbox{(i)} & {\cal L} = - y_N^{}\,\ol{L}_\ell\,\wt{\Phi}\,N_R^{} - \lambda_N^{}\,\phi\,\ol{N_L}\,\psi_R^{} \\
\mbox{(ii)} & {\cal L} = - y_\chi\,\ol{L}_\ell\,\wt{\eta}\,\psi_R^{} - \mu_\phi\,\phi\left(\Phi^\dagger\eta\right) \\
\mbox{(iii)} & {\cal L} = - y_L^{}\,\ol{L}_\ell\,\phi\,L'_R - \lambda_L^{}\,\ol{L'_L}\,\wt{\Phi}\,\psi_R^{} \\
\end{array} \right.
\end{equation}
Here, $N_{L,R}\sim(\1,\1,0)$ is a vectorlike neutral lepton, 
$\eta\sim(\1,\2,1/2)$ is a scalar doublet and 
$L'_{L,R}\sim(\1,\2,-1/2)$ is a vectorlike doublet lepton.
Integrating out these heavy particles, 
we obtain the dim-5 operator eq.\,(\ref{eq:dim5}) in each model. 
The models (i)-(iii) are clearly renormalizable.
These three models have been discussed in the literature. 
The model (i) is equivalent to neutrino portal DM models \cite{Pospelov:2007mp, Batell:2017cmf, McKeen:2018pbb, Blennow:2019fhy}. 
The model (ii) is proposed by ref.\,\cite{Okawa:2020jea} 
as another realization of viable sub-GeV DM coupled mainly to neutrinos. 
The model (iii) is discussed in refs.\,\cite{Restrepo:2015ura, Esch:2016jyx} in conjunction with radiative seesaw scenarios, also known as scotogenic scenarios \cite{Ma:2006km, Bonnet:2012kz, Restrepo:2013aga}. 
In these models, masses of $\psi$ and $\phi$ are free parameters and can be made arbitrarily light.

Alternatively, one can restore the renormalizability by considering that either $\psi$ or $\phi$ (or both) originates in an EW multiplet field.
For example, if $\phi$ is regarded as a neutral component of an EW doublet $\eta$, one can rewrite eq.\,(\ref{eq:Lint_eff}) as
\beq
\ol{\nu_L^{}}_\ell\,\phi\,\psi_R^{} 
    \to \ol{L}_\ell\,\wt{\eta}\,\psi_R^{}\,;
    \quad\eta=\begin{pmatrix} H^+ \\ \frac{1}{\sqrt{2}}\(H+iA\) \end{pmatrix}
\eeq
where $H, A$ and $H^+$ are the neutral and charged components of $\eta$, and hereafter, we consider $H$ to be lighter than $A$ without loss of generality.
This renormalizable extension is known as the lepton portal DM model \cite{Bai:2014osa, Chang:2014tea} (see also ref.\,\cite{Kawamura:2020qxo} for the comprehensive phenomenological analysis). 
If the charged component $H^+$ can be decoupled from the theory while keeping the others light, 
one can effectively obtain eq.\,(\ref{eq:Lint_eff}) at low energies. 
However, in contrast to the neutrino portal DM models, 
the extra scalar masses are related to each other in this model, and 
it is non-trivial whether this model can accommodate a light scalar consistently. 
The general scalar potential in the model is given by 
\begin{align}
V & = m_1^2 (\Phi^\dagger \Phi)^2 + m_2^2 (\eta^\dagger \eta)
    + \lambda_1 (\Phi^\dagger \Phi)^2 
    + \lambda_2 (\eta^\dagger \eta)^2 \nonumber\\
  & \quad 
    + \lambda_3 (\Phi^\dagger \Phi) (\eta^\dagger \eta)
    + \lambda_4 (\Phi^\dagger \eta) (\eta^\dagger \Phi) 
    + \frac{\lambda_5}{2} \left[ (\Phi^\dagger \eta)^2 + {\rm h.c.} \right],
\end{align}
where we imposed a $Z_2$ symmetry: $\eta \to -\eta$ and $\psi \to -\psi$, while all the SM fields are even.
The masses of the extra scalars $H, A$ and $H^+$ are split through the quartic couplilngs to the Higgs doublet $\Phi$, explicitly given by 
\begin{equation}
m_{H^\pm}^2 - m_H^2 = -\frac{(\lambda_4+\lambda_5)v^2}{2} \,,\quad
m_{H^\pm}^2 - m_A^2 = -\frac{(\lambda_4-\lambda_5)v^2}{2} \,.
\end{equation}
Since the LEP bounds require both $H^+$ and $A$ to be heavier than 100\,GeV \cite{Barbieri:2006dq, Pierce:2007ut, Lundstrom:2008ai}, 
a sub-GeV $H$ can be realized only if the quartic couplings are ${\cal O}(1)$. 
As shown in ref.\,\cite{Okawa:2020jea}, 
such a parameter space is still viable and the mass spectrum of our interest can be achieved by tuning three quartic couplings $\lambda_{3,4,5}$ as 
\begin{align}
\lambda_4 + \lambda_5 & \simeq - \frac{2m_{H^\pm}^2}{v^2} \,,\nonumber\\
\lambda_4 - \lambda_5 & \simeq 0 \,,\\
\lambda_3 + \lambda_4 + \lambda_5 & \simeq 0 \,,\nonumber
\end{align}
from which the quartic couplings are sharply predicted as 
\begin{align}
\lambda_3 \simeq - 2 \lambda_4 \simeq -2 \lambda_5 \simeq \frac{2m_{H^+}^2}{v^2} \,.
\end{align}
See refs.\,\cite{Okawa:2020jea, Iguro:2022tmr,Higuchi:2023kbt} for further details of this realization and its experimental constraints. 
In this realization, low-energy DM dynamics is governed mostly by the neutrino portal interaction eq.\,(\ref{eq:Lint_eff}), as the other two scalars are too heavy to affect.
Namely, $H$ can be identified as $\phi$.
In the following sections, we explicitly compute the corrections to the leptonic meson decay within this model. 
The other models (i)-(iii) will predict a very similar result as long as the additional mediators ($N$, $\eta$, $L'$) are heavier than the EW scale.

\section{Corrections to leptonic meson decays}
\label{sec:correction}

In this section, we discuss the corrections to the leptonic decays $M \to \ell \,\ol\nu_\ell$ of a pseudo-scalar meson $M$.
The relevant couplings in our study are given by 
\beq
{\cal L} \supset 
-\frac{\lambda_\ell}{\sqrt{2}}\,\ol{\nu_L^{}}_\ell\,H\,\psi_R^{} 
+\frac{i\lambda_\ell}{\sqrt{2}}\,\ol{\nu_L^{}}_\ell\,A\,\psi_R^{}
+\lambda_\ell\,\ol{\ell_L^{}}\,H^-\,\psi_R^{}+\hc
\label{eq:Lint}
\eeq
with $\ell=e,\mu,\tau$. 
In this work, we assume for simplicity that only one of three couplings $\lambda_\ell$ is non-vanishing. 
This assumption also helps avoid large charged lepton flavor violation (cLFV).

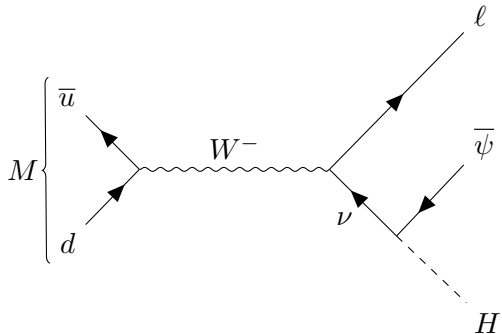
\begin{figure}[t]
    \centering
    \begin{tikzpicture}
        \begin{feynman}
            \vertex (i0);
            \vertex [above left=1cm of i0] (ubar) {$\ol{u}$};
            \vertex [below left=1cm of i0] (d) {$d$};
            \vertex [right=2.5cm of i0] (i1);
            \vertex [above right=2.5cm of i1] (ell) {$\ell$};
            \vertex [below right=1.25cm of i1] (i2);
            \vertex [below right=1.25cm of i2] (H) {$H$};
            \vertex [above right=1.25cm of i2] (psibar) {$\ol\psi$};
            
            \diagram* {
            (ubar) -- [anti fermion] (i0) -- [anti fermion] (d), 
            (i0) -- [boson, edge label=$W^-$] (i1) -- [fermion] (ell),
            (i1) -- [anti fermion, edge label'=$\nu$] (i2) -- [anti fermion] (psibar),
            (i2) -- [scalar] (H),
            };
            
            \draw [decoration={brace}, decorate] (d.south west) -- (ubar.north west) node [pos=0.5, left] {$M$};
            
        \end{feynman}
    \end{tikzpicture}
    \caption{Feynman diagram for the three-body meson decay $M \to \ell\, \ol\psi \,H$}
    \label{fig:three-body}
\end{figure}

As we mentioned above, these interactions modify the $M \to \ell \,\ol\nu_\ell$ observables through two corrections. 
First, eq.\,(\ref{eq:Lint}) induces a new three-body decay $M \to \ell\,\ol\psi\,H$ (fig.\,\ref{fig:three-body}). 
As this is the three-body process with the massive invisible particles, 
the energy distribution of the charged lepton can significantly differ from $M \to \ell\,\ol\nu_\ell$. 
Moreover, the helicity suppression, which is found in the $M \to \ell\,\ol\nu_\ell$ decay, 
is absent in the three-body decay, suggesting a potentially large correction. 
Second, there is a one-loop correction to the weak vertex $W\ell\,\ol\nu_\ell$ (fig.\,\ref{fig:one-loop}).
This correction does not change the lepton energy distribution, 
but enhances or suppresses the partial decay width which in turn 
perturbs the branching fraction of $M \to \ell\,\ol\nu_\ell$ for a particular lepton flavor. 
The most important interaction in our analysis is the first term in eq.\,(\ref{eq:Lint}), which only involves the light scalar and fermion and thus generates the leading corrections. 
However, since we are carrying out the one-loop calculation, 
the second and third terms are necessary to obtain UV-divergence--free results. 
Note that the one-loop corrections also play an important role in canceling an IR divergence stemming from collinear emission of a light $H$ in the three-body process. 

It might be worth mentioning several works that study the leptonic meson decays as a probe of neutrino-philic new physics.
In \cite{Batell:2017cmf}, the three-body Kaon decay process is evaluated in the Dirac neutrino portal DM scenario. 
While the process is completely the same as what we are going to examine in this paper, 
the pion decay and the one-loop corrections are not evaluated. 
In \cite{Berryman:2018ogk}, the impact of a Majoron-like mediator $J$ on the leptonic meson decays is discussed in detail. 
The relevant process is $M \to \ell\,\ol\nu J$ rather than $M\to\ell\,\ol\psi\,H$, and 
the one-loop corrections are not taken into account.
In \cite{Dev:2024ygx}, the one-loop vertex correction is evaluated for the Majoron-like mediator, and its role in canceling the IR divergence is confirmed for the first time. 
The work of \cite{Dev:2024ygx} is followed up in \cite{Dev:2025tdv}, where several types of neutrino-philic dark sectors, including neutrino-portal-like models, are systematically studied. 
They include both three-body and one-loop corrections, although all calculation is carried out only using effective DM-neutrino interactions like eq.\,(\ref{eq:Lint_eff}) (or equally the first term of eq.\,(\ref{eq:Lint})).
In this case, a UV divergent term appears in the one-loop calculation, 
and it is subtracted in a minimal way in refs.\,\cite{Dev:2024ygx, Dev:2025tdv}.
This approach might be fairly reasonable, because the leading contribution arises mainly from the light particles as we will see in Sec.\,\ref{sec:analysis}. 
In this paper, we keep the second and third terms in eq.\,(\ref{eq:Lint}) to maintain the renormalizability and predictability, 
so that our calculation does not lead to any divergence in physical observables. 

In the rest of this section, we explicitly calculate the three-body and one-loop corrections to the meson decays with some technical details. 
Readers who are only interested in the comparison with experimental data can skip over it to Sec.\,\ref{sec:analysis}.

\subsection{Three-body decay $M \to \ell\,\ol\psi\,H$}

We consider the decay of a pseudo-scalar meson $M$ that is composed of $q$ and $q^\prime$ quarks.
With eq.\,(\ref{eq:Lint}), the amplitude square for $M \to \ell\,\ol\psi\,H$ is calculated as 
\begin{align}
|{\cal M}(M \to \ell\,\ol\psi\,H)|^2 
    & = 2 |V_{qq'}|^2 |\lambda_\ell|^2 G_F^2 f_M^2 \(\frac{1}{q^2}\)^2 \nonumber\\
    & \quad \times \left\{ q^4 (p_\ell^{}\cdot p_\psi^{}) + m_\ell^2 \[ 2(q\cdot p_\psi^{})(q^2+ q\cdot p_\ell^{}) - q^2 (p_\ell^{}\cdot p_\psi^{}) \] \right\}\,, \label{eq:Msq1} 
\end{align}
where $V$ is the Cabibbo-Kobayashi-Maskawa (CKM) matrix, $f_M$ is the decay constant of $M$, and 
$q$ is the momentum of the neutrino propagator. 
Using $q^2=(p_\psi^{}+p_H^{})^2$ and $p^2=(p_H^{}+p_\ell^{})^2$ and 
\begin{align}
p_\ell^{} \cdot p_\psi^{} & = -\frac{1}{2} \(q^2+p^2-m_M^2-m_H^2\) \,,\\
q \cdot p_\psi^{} & = \frac{1}{2} \(q^2+m_\psi^2-m_H^2\) \,,\\
q \cdot p_\ell & = -\frac{1}{2} \(q^2-m_M^2+m_\ell^2\) \,,
\end{align}
$|\M|^2$ can be expressed only by $q^2$ and $p^2$.
Note that the $p^2$ dependence of $|\M|^2$ comes only from the momentum product $p_\ell^{}\cdot p_\psi^{}$.

The differential decay width is given by 
\beq
\d\Gamma(M \to \ell\,\ol\psi \,H) = \frac{|\M|^2}{256\pi^3 m_M^3} \, \d p^2 \, \d q^2 \,.
\eeq
The decay width is obtained by performing the integrals over
\beq
p_{\rm min}^2 \leq p^2 \leq p_{\rm max}^2 \,,\quad
(m_\psi^{}+m_H^{})^2 \leq q^2 \leq (m_M^{}-m_\ell^{})^2 \,,
\eeq
where 
\beq
\left\{\begin{matrix} p_{\rm max}^2 \\ p_{\rm min}^2 \end{matrix} \right\}
= \(E_H^*+E_\ell^*\)^2-\(\sqrt{(E_H^*)^2-m_H^2}\mp\sqrt{(E_\ell^*)^2-m_\ell^2}\)^2
\eeq
with 
\beq
E_H^* = \frac{q^2-m_\psi^2+m_H^2}{2|q^2|^{1/2}} \,,\quad
E_\ell^* = \frac{m_M^2-q^2-m_\ell^2}{2|q^2|^{1/2}} \,.
\eeq
The integral over $p^2$ can be easily performed, yielding the differential rate, 
\begin{align}
\frac{\d\Gamma(M \to \ell\,\ol\psi \,H)}{\d q^2} 
    &=\frac{|V_{qq'}|^2|\lambda_\ell|^2G_F^2f_M^2}{512\pi^3 m_M^3} 
        \frac{\sqrt{\lambda(q^2,m_\psi^2,m_H^2)\,\lambda(m_M^2,q^2,m_\ell^2)}}{q^6} \,g(q^2) \,,\label{eq:dGamma/dq2}
\end{align}
where 
\begin{align}
g(q^2)&=\(q^2-m_H^2+m_\psi^2\) \left\{m_M^2\(q^2+m_\ell^2\)-\(q^2-m_\ell^2\)^2\right\} \,.
\end{align}
The equation (\ref{eq:dGamma/dq2}) provides the distribution of the missing mass square $q^2=(p_M^{}-p_\ell^{})^2$, 
which is measured in search for the $K\to\ell\,\ol\nu_\ell$ decay in the NA62 experiment \cite{NA62:2011aa, NA62:2012lny}. 
Contrarily, the energy distribution of the final-state charged lepton is measured in other experiments like the PIENU collaboration \cite{PiENu:2015seu}, where the parent meson particle is at rest in the lab-frame. 
In the latter case, the differential decay width with respect to $E_\ell$ might be more useful:
\begin{align}
\frac{\d\Gamma(M \to \ell\,\ol\psi \,H)}{\d E_\ell} 
    &=2m_M\left.\frac{\d\Gamma(M \to \ell\,\ol\psi\, H)}{\d q^2}\right|_{q^2=m_M^2+m_\ell^2-2m_M^{}E_\ell}
\end{align}
Here, the integration range is given by 
\beq
m_\ell^{} \leq E_\ell \leq \frac{m_M^2+m_\ell^2-(m_\psi^{}+m_H^{})^2}{2m_M^{}} \,.
\eeq

%====================================
%=====================================

The total decay width is obtained by performing the $q^2$-integral,
\beq
\Gamma(M\to\ell\,\ol\psi\, H) = 
\int_{(m_\psi^{}+m_H^{})^2}^{(m_M^{}-m_\ell^{})^2} \d q^2 \frac{\d\Gamma}{\d q^2} 
= \frac{|V_{qq'}|^2|\lambda_\ell|^2G_F^2m_M^3f_M^2}{512\pi^3}\,f(x_\ell,x_\psi,x_H^{}) \,.
\eeq
where $x_i=m_i^2/m_M^2$ and $f$ is a dimensionless function, given by 
\beq
f(x_\ell^{},x_\psi^{},x_H^{})=
\frac{1}{m_M^6} \int \d q^2 
\frac{\sqrt{\lambda(q^2,m_\psi^2,m_H^2)\,\lambda(m_M^2,q^2,m_\ell^2)}}{q^6} \,g(q^2) \,.
\eeq
This integral is numerically performed in our analysis. 
In the limit of $x_\psi\to0$, we can analytically perform the integral, finding \footnote{The equation (\ref{eq:f1approx}) completely agrees with eq.\,(B3) of \cite{Dev:2024ygx} except for 
the first term of the third line and the argument of the second arctanh function. 
Although we were not able to identify the source of this discrepancy, 
we confirmed eq.\,(\ref{eq:f1approx}) does reproduce the limiting expression eq.\,(B5) of that paper, 
while eq.\,(B3) does not lead to eq.\,(B5).
There might be typos in ref.\,\cite{Dev:2024ygx}.}
\begin{align}
f(x_\ell,x_\psi,x_H^{}) 
    &\simeq\frac{\sqrt{\lambda(1,x_\ell,x_H^{})}}{3(1-x_\ell)}
    \(1+10x_H^{}+x_H^2-x_\ell(3-x_H^{})^2+x_\ell^2(18-19x_H^{})-10x_\ell^3\) \nonumber\\
    &\quad+\(2x_H^{}(1+x_H^{})-x_\ell(1-3x_H^2)-2 x_\ell^2(1-3x_H^{})+x_\ell^3 \) 
        \arctanh \frac{\sqrt{\lambda(1,x_\ell,x_H^{})}}{1-x_H^{}+x_\ell} \nonumber\\
    &\quad-\(2x_H^{}(1-3x_\ell^2)-x_\ell(1-x_\ell)^2+\frac{(2-x_\ell+4x_\ell^2-3x_\ell^3)x_H^2}{(1-x_\ell)^2}\) \nonumber\\
    &\quad\times \arctanh\frac{(1-x_\ell)\sqrt{\lambda(1,x_\ell,x_H^{})}}{(1-x_\ell)^2-x_\ell(1+x_H^{})} \,.
\label{eq:f1approx}
\end{align}
Taking $m_H^{} \ll 1$, we can further simplify $f(x_\ell, x_\psi, x_H)$:
\beq
f(x_\ell,x_\psi,x_H) \simeq \frac{1}{3}\(1-9x_\ell+18x_\ell^2-10x_\ell^3+3x_\ell^2(2-x_\ell)\log x_\ell+3x_\ell(1-x_\ell)^2\log\frac{(1-x_\ell)^2}{x_\phi}\) \,.
\eeq
which agrees with eq.\,(B5) of \cite{Dev:2024ygx}, given $\arctanh(\frac{1-x_\ell}{1+x_\ell})=-\frac{1}{2}\log x_\ell$. 

We shall estimate the expected impact of the three-body decay for understanding.
Using the tree level $M\to\ell\,\ol\nu_\ell$ decay width in the SM, 
\beq
\Gamma_0 \left(M \to \ell \,\ol\nu_\ell \right) = \frac{|V_{qq^\prime}|^2G_F^2 f_M^2 m_\ell^2 m_M^{}}{8\pi} \left(1-\frac{m_\ell^2}{m_M^2}\right)^2 \,,
\label{eq:Gamma_SM}
\eeq
we obtain the three-body to two-body decay width ratio, 
\beq
\frac{\Gamma(M\to\ell\,\ol\psi \,H)}{\Gamma_0(M\to\ell\,\ol\nu_\ell)} 
\simeq \frac{|\lambda_\ell|^2}{64\pi^2} \frac{m_M^2}{m_\ell^2} f(x_\ell^{},x_\psi^{},x_H^{}) \,.
\eeq
The phase-space factor $\sim1/(64\pi^2)$ can be compensated by the meson-to-lepton mass ratio $m_M^2/m_\ell^2$, which is large especially in the $\ell=e$ case.
Given the current level of precision is ${\cal O}(0.1-1)\%$, 
the measurements of the leptonic meson decays have the potential to probe the relevant coupling $|\lambda_\ell|$ down to ${\cal O}(0.01)$ for $\ell=e$ and ${\cal O}(1)$ for $\ell=\mu$. 

It should be noted that there is an additional contribution to the three-body process $M \to \ell\, \ol\psi\, H$ from $M \to (H^-)^* H \to \ell\,\ol\psi\, H$. 
Here, $(H^+)^*$ denotes an off-shell charged scalar.
This process carries the same coupling scaling as the process we discussed above, 
but receives a suppression from the heavy charged scalar mass. 
Thus we can ignore the charged scalar mediated process.

\subsection{One-loop correction to $M \to \ell \,\overline{\nu}_\ell$}

%====================================
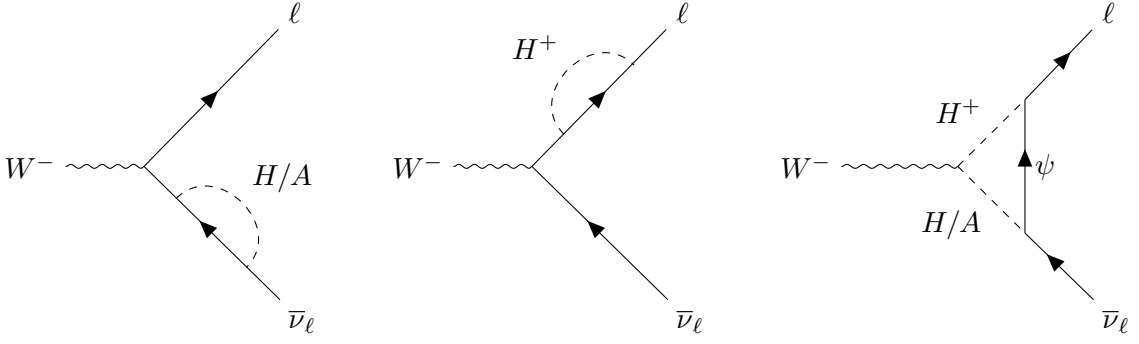
\begin{figure}[t]
    \centering
    \begin{tikzpicture}
        \begin{feynman}
            \vertex (i0) {$W^-$};
            \vertex [right=1.5cm of i0] (i1);
            \vertex [above right=2.5cm of i1] (ell) {$\ell$};
            \vertex [below right=2.5cm of i1] (nubar) {$\ol{\nu}_\ell$};
            \vertex [below right=0.6cm of i1] (i2);
            \vertex [below right=1.3cm of i2] (i3);
            
            \diagram* {
            (i0) -- [boson] (i1) -- [fermion] (ell),
            (i1) -- [anti fermion, edge label'] (nubar),
            (i2) -- [scalar, edge label=$H/A$, half left] (i3),
            };
            
        \end{feynman}
    \end{tikzpicture}
    \hspace{0.5cm}
    \begin{tikzpicture}
        \begin{feynman}
            \vertex (i0) {$W^-$};
            \vertex [right=1.5cm of i0] (i1);
            \vertex [above right=2.5cm of i1] (ellbar) {$\ell$};
            \vertex [below right=2.5cm of i1] (nubar) {$\ol{\nu}_\ell$};
            \vertex [above right=0.6cm of i1] (i2);
            \vertex [above right=1.3cm of i2] (i3);
            
            \diagram* {
            (i0) -- [boson] (i1) -- [fermion] (ell),
            (i1) -- [anti fermion] (nubar),
            (i2) -- [scalar, edge label=$H^+$, half left] (i3),
            };
                        
        \end{feynman}
    \end{tikzpicture}
    \hspace{0.5cm}    
    \begin{tikzpicture}
        \begin{feynman}
            \vertex (i0) {$W^-$};
            \vertex [right=2cm of i0] (i1);
            \vertex [above right=2.5cm of i1] (ell) {$\ell$};
            \vertex [below right=2.5cm of i1] (nubar) {$\ol{\nu}_\ell$};
            \vertex [above right=1.25cm of i1] (i2);
            \vertex [below right=1.25cm of i1] (i3);
            
            \diagram* {
            (i0) -- [boson] (i1),
            (i1) -- [scalar, edge label=$H^+$] (i2) ,
            (i1) -- [scalar, edge label'=$H/A$] (i3),
            (ell) -- [anti fermion] (i2) -- [anti fermion, edge label=$\psi$] (i3) -- [anti fermion] (nubar),
            };
                        
        \end{feynman}
    \end{tikzpicture}
    \caption{
    One-loop corrections to the weak-vertex $W \ell\,\ol\nu_\ell$ in our model. 
    The decay of a meson $M$ is induced from attaching the $W$ boson to $\ol{u}$ and $d$ external lines like in fig.\ref{fig:three-body}.
    }
    \label{fig:one-loop}
\end{figure}
%=====================================

Next, we study the one-loop weak-vertex corrections.
The relevant diagrams are shown in fig.\,\ref{fig:one-loop}. 
In general, the vertex corrections are expressed in the form, 
\beq
{\cal L}^W_{\rm eff}&= \frac{g}{\sqrt{2}}  W_\mu (q) \left \{ \delta g_W^{} \,    \overline{\ell_L} (p) \gamma^\mu \nu_L(p-q)+ f_{W}^{\mu} \, m_{\ell} \,  \overline{\ell_R} (p) \nu_L (p-q)\right \},
\eeq
where $\delta g_W^{}$ and $f^\mu_{W}$ are respectively given by 
\beq
\delta g_W^{}  = \frac{|\lambda_\ell|^2}{32 \pi^2} \int^1_0 dx \int^1_0 dy & \Bigg [ \, \frac{y}{2}  \log \left ( \frac{ y m^2_H +(1-y) m_N^2 }{\Delta (m_{H_\pm},m_A,m_\ell)}  \right ) +\frac{y}{2}  \log \left ( \frac{ y m^2_A +(1-y) m_N^2 }{\Delta (m_{H_\pm},m_A,m_\ell)}\right )  \nonumber \\
&+ y  \log \left ( \frac{ y m^2_{H_\pm} +(1-y) m_N^2 - m^2_\ell y (1-y) }{\Delta (m_{H_\pm},m_H,m_\ell)}\right ) \Bigg ], 
\eeq
and 
\beq
f^\mu_{W}=& \frac{|\lambda_\ell|^2}{16 \pi^2} \int^1_0 dx \int^1_0 dy \, \frac{y}{2} \, \left ( \frac{1}{\Delta (m_{H_\pm},m_H,m_\ell)} +\frac{1}{\Delta (m_{H_\pm},m_A,m_\ell)} \right )   \nonumber \\
&\times  y(x-1) \left \{ 2 p^\mu (1-y) +q^\mu (1+2 xy) \right \}  \nonumber \\
\equiv & f_p \,p^\mu + f_q \,q^\mu .
\eeq
Here, $\Delta (m_{H_\pm},m_S,m_\ell)$ is given by
\beq
\Delta (m_{H_\pm},m_S,m_\ell)=&m^2_{H_\pm}y (1-x) +(m^2_N- m^2_\ell) (1-y) + (m^2_S -q^2 )xy \nonumber \\
&+x^2y^2q^2 +(1-y)^2 m^2_{\ell}+xy (1-y) (m^2_\ell+q^2).
\eeq

Taking into account the one-loop weak-vertex corrections, 
the amplitude square for the leptonic meson decay is modified as
\beq
|{\cal M}(M\to\ell\,\ol{\nu}_\ell)|^2& \simeq \frac{|V_{qq^\prime}|^2g^4f_M^2}{16 m^2_W} \left \{ m^2_\ell \left (m_M^2- m^2_\ell \right ) \right \} \times \Delta^\ell_W \,.
\eeq
The new physics contribution is encoded in $\Delta_W^\ell$:
\beq
\Delta^\ell_W = \left (1+ \delta g_W^{} \right )^2 +2\left (1+ \delta g_W^{} \right ) \left \{ \left ( \frac{1}{2} f_p +f_q \right ) m^2_M+f_p \frac{m^2_\ell}{2} \right \} + \left \{ \left ( \frac{1}{2} f_p +f_q \right ) m^2_M+f_p \frac{m^2_\ell}{2} \right \}^2.
\eeq
We have $\Delta_W^\ell=1$ in the absence of the new physics contribution.
Then, the decay width is simply modified as 
\beq
\Gamma \left(M\to\ell\bar\nu_\ell\right)=\Delta^\ell_W\,\Gamma_0 \left(M\to\ell\bar\nu_\ell\right)\,,
\eeq
where $\Gamma_0 \left(M\to\ell\,\ol\nu_\ell\right)$ is the SM prediction at the tree level, see eq.\,(\ref{eq:Gamma_SM}).

The leading correction in $\Delta_W^\ell$ arises from the first term, $\left(1+\delta g_W^{}\right)^2$.
When the extra scalars $H^\pm$ and $A$ are much heavier than the other particles in the loop, $\delta g_W^{}$ is approximated to 
\beq
\label{eq;gWapp}
\delta g_W^{} \simeq -\frac{|\lambda_\ell|^2}{64\pi^2} \log\left(\frac{m_{H_\pm}^{}}{m_H^{}}\right) < 0\,.
\eeq
$\delta g_W^{}$ is negative and has the magnitude of ${\cal O}(0.01) \times |\lambda_\ell|^2$ for $\frac{m_{H_\pm}^{}}{m_H^{}}={\cal O}(100)$, 
meaning that the one-loop vertex correction modifies ${\rm BR}(M\to\ell\,\ol\nu_\ell)$ by only $1\,\%$ or smaller for a perturbative coupling $\lambda_\ell$.
While it is a quite small correction, the $M\to\ell\,\ol\nu_\ell$ observables are experimentally measured with a precision less than 1\,\%. 
Thus we will obtain a non-trivial bound on $|\lambda_\ell|$ of ${\cal O}(1)$ or smaller.

%%%%%%%%%%%%%%%%%%%%%%%%%%%%%%%%%%%%%%%%%%%%%%%%%%%%%%%%%%%
\begin{table}[t]
\begin{center}
\begin{tabular}{|cc|}
\hline
Observables & Values  \\ \hline
$m_e$ & 0.51100 MeV  \\
 $m_\mu$ & 105.66 MeV \\
%$m_\tau$ & 1.777 GeV  && 
 $m_\pi$ & 139.57 MeV \\
  $m_K^{}$ & 493.68 MeV 
 \\ \hline
\end{tabular}
\end{center}
\caption{The values used in our analysis \cite{ParticleDataGroup:2024cfk}.}
\label{tab:parameter}
\end{table}
%%%%%%%%%%%%%%%%%%%%%%%%%%%%%%%%%%%%%%%%%%%

\section{Analysis}
\label{sec:analysis}

Now we are ready to compare our calculation with the experimental results.
In this work, we focus particularly on the lepton flavor universality of meson decays, i.e. 
the ratio of ${\rm BR}(M\to e \,\ol{\nu})$ to ${\rm BR}(M\to \mu \,\ol{\nu})$:
\beq
R^{(M)}_{e/\mu}= \frac{{\rm BR}(M\to e\,\ol{\nu})}{ {\rm BR}(M\to \mu\,\ol{\nu})}.
\eeq
The theoretical uncertainty from the CKM matrix elements, the meson decay constants, and the determination of the EW observables is significantly canceled in this quantity, allowing us to carry out very clean and robust comparison.
The SM predictions have been studied with a great precision in refs.\,\cite{Finkemeier:1995gi,Cirigliano:2007xi}, where 
radiative corrections up to ${\cal O}(e^2p^4)$ in Chiral Perturbation Theory (ChPT) are taken into account. 
If the radiative corrections to the leptonic meson decay width are the same as in the SM, 
$R^{(M)}_{e/\mu}$ can be expressed in our model as
\beq
\label{eq;Remu}
R^{(M)}_{e/\mu}= R^{(M),\,\SM}_{e/\mu} \left( \frac{\Delta^{e}_{W}+B^{(M)}_e}{\Delta^{\mu}_{W}+B^{(M)}_\mu} \right)\,,
\eeq
where $B^{(M)}_\ell$ is defined as
\beq
B^{(M)}_\ell = \frac{\Gamma \left ( M \to \ell \, \ol\psi \, H \right )}{\Gamma_0 \left ( M \to \ell \, \ol{\nu} \right )}.
\eeq
Here, $R^{(M), \, \SM}_{e/\mu}$ denotes the SM predictions including the radiative corrections.

In the Kaon case ($M=K$), the SM prediction is \cite{Cirigliano:2007xi}
\beq
R^{(K), \,\SM}_{e/\mu}=\left ( 2.477 \pm 0.001 \right ) \times 10^{-5},
\eeq
while the experimental average is \cite{ParticleDataGroup:2024cfk} 
\beq
R^{(K), \,{\rm exp}}_{e/\mu}=(2.488\pm0.009)\times 10^{-5}.
\eeq
The SM prediction is slightly below the experimental average.
In the pion case ($M=\pi$), the SM prediction is \cite{Cirigliano:2007xi}
\beq
R^{(\pi), \,\SM}_{e/\mu}=\left ( 1.2352 \pm 0.0001 \right ) \times 10^{-4},
\eeq
while the experimental average is \cite{ParticleDataGroup:2024cfk} 
\beq
R^{(\pi), \,{\rm exp}}_{e/\mu}=(1.2327\pm0.0023)\times 10^{-4}.
\eeq
The SM prediction is slightly above the experimental average.

In the following, we require $R^{(M)}_{e/\mu}$ to lie within $2\sigma$ of the experimental average values 
to derive the constraints from the $K$ and $\pi$ decays.
As we mentioned in the beginning of Sec.\,\ref{sec:correction}, 
we turn on only one of three couplings $\lambda_\ell$ ($\ell=e,\mu,\tau$) to avoid the strong cLFV bounds. 
In particular, the electro-philic ($\lambda_e\neq0$) or muon-philic ($\lambda_\mu\neq0$) case is relevant for the purpose of this work, 
and we will discuss two cases below. 
The case of the tau coupling will be studied elsewhere.

\subsection{Electron flavor}

%%%%%%%%%%%%%%%%%%%%%%%%%%%%%%%%%%
\begin{figure}[t]
\begin{center}
\includegraphics[width=7cm]{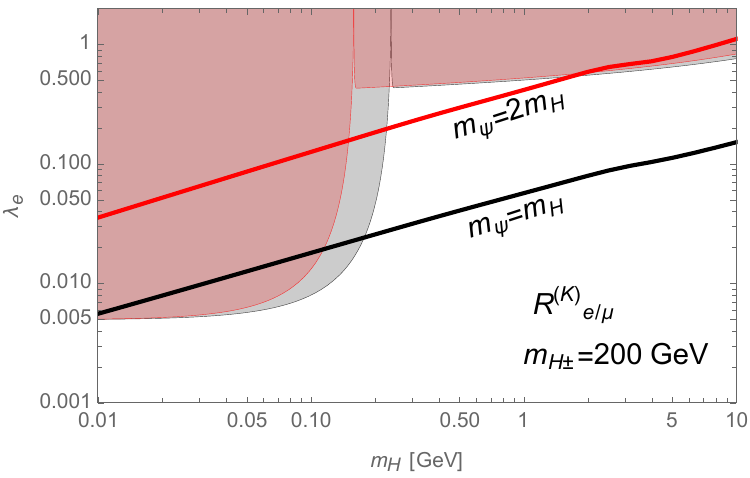}
\includegraphics[width=7cm]{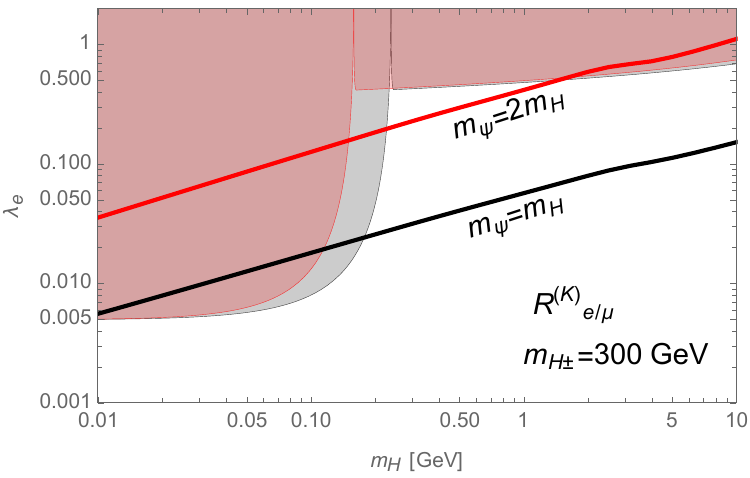}

\vspace{0.3cm}

\includegraphics[width=7cm]{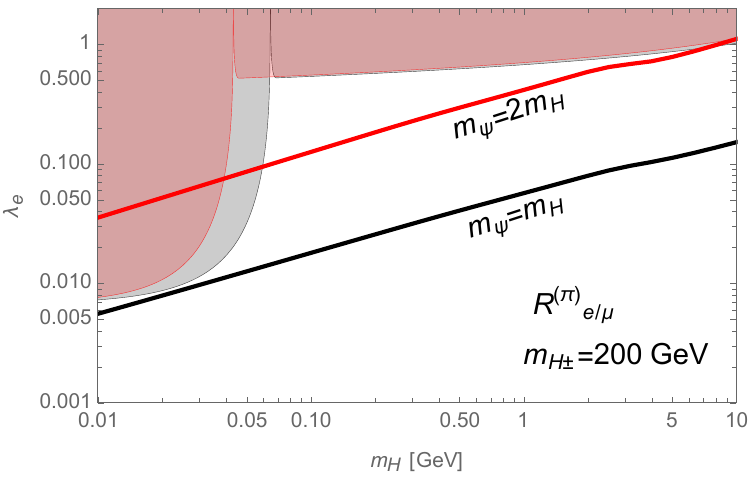}
\includegraphics[width=7cm]{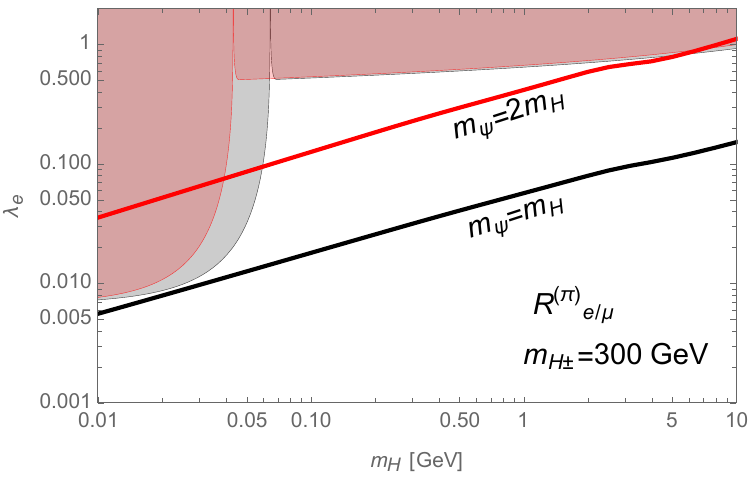}
\caption{
The constraints from the lepton universality measurements of $K\to\ell\bar\nu$ (top) and $\pi\to\ell\bar\nu$ (bottom) 
in the electro-philic case, where $\lambda_e\neq0$ and $\lambda_\mu=\lambda_\tau=0$. 
The $2\sigma$ exclusion regions are shaded with $m_\psi=m_H$ (gray) and $m_\psi=2m_H$ (red). 
The solid lines correspond to the predictions in the thermal freeze-out scenario with $m_H=m_\psi$ (gray) and $m_H=2 m_\psi$ (red). 
The thermal relic abundance of the DM candidate agrees with the Planck observation $\Omega h^2=0.12$ on those lines.
\label{fig:mDMvsYukawa_E}
}
\end{center}
\end{figure}
%%%%%%%%%%%%%%%%%%%%%%%%%%%%%%%%%%

First, we discuss the electro-philic case, where $\psi$ couples exclusively to the electron flavor, 
i.e. $\lambda_e\neq0$ and $\lambda_\mu=\lambda_\tau=0$.
In this case, $R^{(M)}_{e/\mu}$ is expressed as
\beq
\label{eq;Remu_e}
R^{(M)}_{e/\mu}= R^{(M),\,\SM}_{e/\mu} \left(\Delta^{e}_{W} + B^{(M)}_e\right)\,.
\eeq
Here, $B^{(M)}_e$ is non-vanishing only when $m_\psi +m_H < \sqrt{m^2_M +m^2_e}$.

In fig.\,\ref{fig:mDMvsYukawa_E}, 
we show the constraints on $|\lambda_e|$ from the lepton flavor universality of $K\to\ell\,\ol\nu$ (top) and $\pi\to\ell\,\ol\nu$ (bottom). 
The charged scalar mass is chosen to be $m_{H_\pm}=200\,\GeV$ on the left panels and $m_{H_\pm}=300\,\GeV$ on the right panels, but it is easily observed that the effects of the charged scalar mass are limited. 
The shaded regions are excluded by the measurements of 
$R^{(K)}_{e/\mu}$ and $R^{(\pi)}_{e/\mu}$ for $m_\psi=m_H$ (gray) and $m_\psi=2m_H$ (red). 
Below the kinematical thresholds $m_\psi +m_H < \sqrt{m^2_M +m^2_e}$, 
the three-body contribution $B^{(M)}_e$ considerably enhances $R^{(M)}_{e/\mu}$, 
due to the absence of the helicity suppression, 
thereby offering a strong upper bound $|\lambda_e| \lesssim {\cal O}(0.01)$.
In this mass region, the one-loop contribution is negligibly small. 
Above the thresholds $m_\psi +m_H > \sqrt{m^2_M +m^2_e}$, 
only the one-loop correction $\Delta^e_W$ can contribute to $R^{(K)}_{e/\mu}$. 
The bound in this mass region is almost independent of $m_H^{}$ and one can obtain a weaker bound $|\lambda_e|\lesssim0.5$. 
For comparison, the solid lines show the predictions in the thermal freeze-out scenario with $m_\psi=m_H$ (gray) and $m_\psi=2 \, m_H$ (red). 
The observed DM abundance can be achieved on the lines. 
The meson decay bounds cover the mass region below ${\cal O}(100)\,\MeV$, which is otherwise unconstrained by the existing experimental search \cite{Okawa:2020jea}. 

One may find a general tendency that $R^{(K)}_{e/\mu}$ gives a stronger bound than $R^{(\pi)}_{e/\mu}$ over the entire mass region, despite the smaller error in $R^{(\pi), \,{\rm exp}}_{e/\mu}$ than in $R^{(K), \,{\rm exp}}_{e/\mu}$. 
In the three-body dominant mass region, 
this is understood from the fact that the phase-space is narrower and 
the three-body contribution $B_e^M\propto m_M^2/m_e^2$ is smaller in the pion decay than in the Kaon decay.
In the one-loop dominant mass region, this is due mostly to the sign of $\delta g_W$ and the current experimental status of $R^{(M)}_{e/\mu}$.
As shown in eq.\,(\ref{eq;gWapp}), 
$\delta g_W$ is always negative and reduces $R^{(M)}_{e/\mu}$ from the SM values in our parameter region.
In the Kaon case, the SM prediction is about $1\sigma$ below the experimental average $R^{(K), \,{\rm exp}}_{e/\mu}$ and the new physics contribution shifts $R^{(K)}_{e/\mu}$ further away from the experimental value.
In the pion case, the SM prediction is about $1\sigma$ above the experimental average. 
Thus the $\delta g_W$ contribution first compensates the deviation, making $R^{(\pi)}_{e/\mu}$ closer to the experimental value, and then deviates it from the experiment one as $\lambda_\mu$ becomes larger. 
The resulting constraint from the pion decay is a little weaker than the Kaon decay.

Note that we only use the total branching ratios in our analysis.
In the actual experimental analysis, however, the energy distribution of the final-state charged lepton, or equally the distribution of the missing mass $m_{\rm miss}^2=(p_M^{}-p_\ell)^2$, provides more useful information. 
In particular, the signal events from $M\to\ell\,\ol{\nu}_\ell$ exhibit a sharp peak around $m_{\rm miss}^2=0$, which is utilized to separate the signals from the backgrounds.
Given that the three-body process $M\to\ell\,\ol\psi\,H$ involves the massive invisible particles in the final state, it can significantly disturb the charged lepton distribution, extending it to the large $m_{\rm miss}^2$ region. 
Search for such a pronounced event shape in the high-$m_{\rm miss}^2$ tail region 
would improve the sensitivity to $|\lambda_e|$. 
See refs.\,\cite{NA62:2017ynf, NA62:2020mcv, NA62:2021bji, NA62:2025csa, PIENU:2021clt} 
for analogous experimental search for $K$ and $\pi$ decays to a charged lepton and an invisible particle, $M \to \ell N$ and $M \to \ell \bar\nu X$, 
where $N$ denotes a heavy neutral lepton and $X$ a bosonic invisible particle. 
To perform a similar spectral analysis, additional information about the acceptance and the SM backgrounds are needed, in addition to the theoretical calculation of the distribution. 
The dedicated spectral analysis will be left for future work. 
Our analysis is thus expected to give conservative bounds.

\subsection{Muon flavor}

%%%%%%%%%%%%%%%%%%%%%%%%%%%%%%%%%%
\begin{figure}[t]
\begin{center}
\includegraphics[width=7cm]{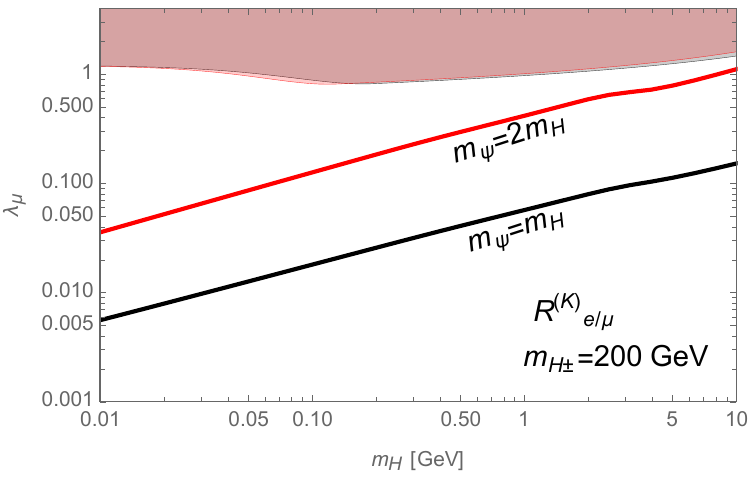}
\includegraphics[width=7cm]{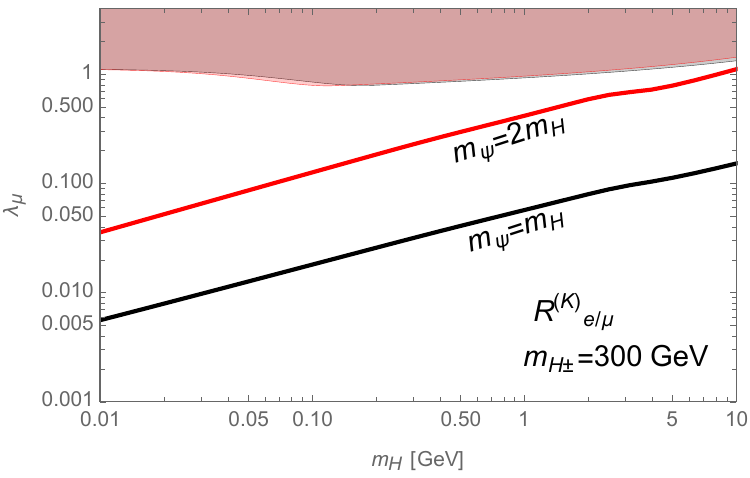}

\vspace{0.3cm}

\includegraphics[width=7cm]{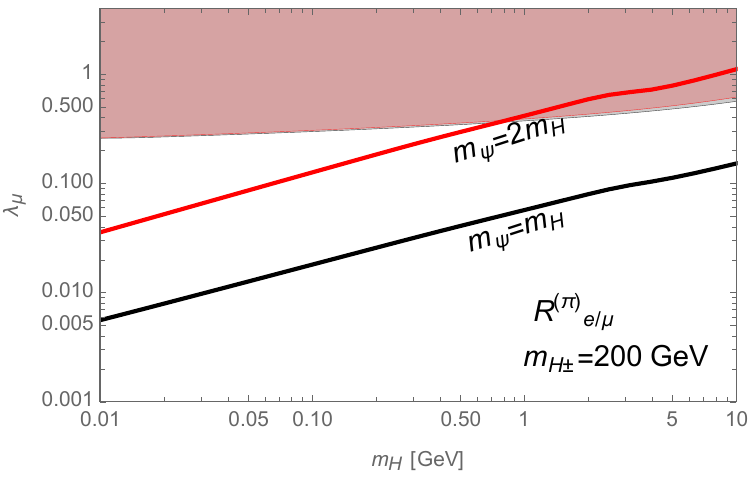}
\includegraphics[width=7cm]{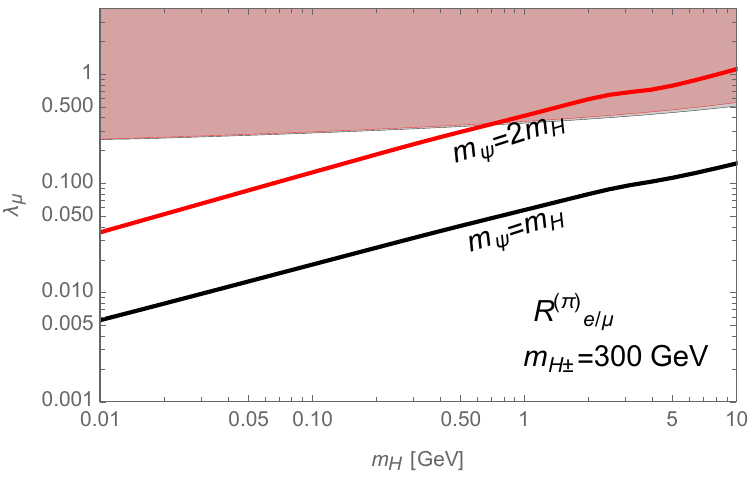}
\caption{
The constraints from the lepton universality measurements of $K\to\ell\bar{\nu}$ (top) and $\pi\to\ell\bar\nu$ (bottom) 
in the muon-philic case, where $\lambda_\mu\neq0$ and $\lambda_e=\lambda_\tau=0$.
The $2\sigma$ exclusion regions are shaded with $m_\psi=m_H$ (gray) and $m_\psi=2m_H$ (red). 
The two regions almost overlap.
The solid lines correspond to the predictions for the thermal freeze-out production 
with $m_H=m_\psi$ (gray) and $m_H=2 m_\psi$ (red). 
The thermal relic abundance of the DM candidate agrees with the Planck observation $\Omega h^2=0.12$ on those lines.
\label{fig:mDMvsYukawa_Muon}
} 
\end{center}
\end{figure}
%%%%%%%%%%%%%%%%%%%%%%%%%%%%%%%%%%

Next, we consider the muon-philic case, where $\lambda_\mu\neq0$ and $\lambda_e=\lambda_\tau=0$.
In this case, only $M\to\mu\,\ol{\nu}_\mu$ is modified, and $R^{(M)}_{e/\mu}$ is given by
\beq
\label{eq;Remu_mu}
R^{(M)}_{e/\mu}=  \frac{R^{(M),\,\SM}_{e/\mu}}{\Delta^{\mu}_{W}+B^{(M)}_\mu} \,.
\eeq
In fig.\,\ref{fig:mDMvsYukawa_Muon}, 
we show the constraints on $|\lambda_\mu|$ from the lepton flavor universality of $K\to\ell\,\ol\nu$ (top) and $\pi\to\ell\,\ol\nu$ (bottom). 
The charged scalar mass is chosen to be $m_{H_\pm}=200\,\GeV$ on the left panels and $m_{H_\pm}=300\,\GeV$ on the right panels, but the results are not sensitive to the mass scale.
The shaded regions are excluded by the measurements of $R^{(K)}_{e/\mu}$ and $R^{(K)}_{e/\mu}$ for $m_\psi=m_H$ (gray) and $m_\psi=2 \, m_H$ (red). The two regions almost overlap.
Compared to the electro-philic case, 
the phase-space of the three-body decay is narrower and the helicity suppression of $M\to\mu\,\ol\nu_\mu$ is milder. 
For these reasons, the one-loop correction $\Delta^{\mu}_{W}$ forms the leading contribution to $R^{(M)}_{e/\mu}$ even if $B^{(M)}_\mu$ is non-vanishing. 
The resultant upper bound on $|\lambda_\mu|$ is much weaker and at most 0.3 across a broad mass region.
One can also see that $R^{(\pi)}_{e/\mu}$ gives a stronger bound than $R^{(K)}_{e/\mu}$. 
This is due to the fact that the one-loop corrections always decreases $\Delta^{\mu}_{W}$ and enhances $R^{(M)}_{e/\mu}$, which is opposite to the electro-philic case.

\section{Other constraints}
\label{sec:others}

The neutrino portal interactions can generally modify all decay channels that involve neutrinos in their final states. 
Among the most interesting channels are the muon Michel decay, tau decay, and $Z$ invisible decay. 
In particular, the muon Michel decay and $Z$ invisible decay are precisely measured and 
might provide a more powerful probe, especially to the high mass region in the $Z$ decay case.
On the other hand, the absence of the helicity suppression in these decay modes would weaken the significance of new physics corrections.
Here we briefly discuss the bounds from the modification of the invisible $Z$ decay width $\Gamma (Z \to {\rm inv})$.

In analogy to the correction of the leptonic meson decays, 
the neutrino portal interactions can induce new invisible decay modes, $Z \to \ol\nu\,\psi\,H$ and $Z \to \ol\psi\,\nu\,H$, and generate one-loop corrections to the $Z$ boson coupling to neutrinos, both of which can modify the invisible $Z$ decay width. 
Moreover, the $Z$ boson can decay into $\overline{\psi} \psi$ via one-loop diagrams, which also makes additional contribution to $\Gamma(Z \to {\rm inv})$.
This latter process is, however, proportional to the fourth power of $\lambda_{e, \mu}$, while the one-loop corrections to $Z\to\nu\ol\nu$ and the three-body processes only carry the second power of $\lambda_{e, \mu}$. 
Thus, the $Z\to\ol\psi\psi$ contribution is relatively small as far as $|\lambda_{e, \mu}|<1$. 
Using the experimental value $\Gamma (Z \to {\rm inv})_{\rm exp}=499.3 \pm 1.5$\,MeV and 
the SM prediction $\Gamma (Z \to {\rm inv})_{\rm SM}=501.435 \pm 0.045$ MeV \cite{ParticleDataGroup:2024cfk}, 
one can evaluate the upper bound $|\lambda_{e,\mu}|<1.22$ for $m_A=200\,\GeV$ and $m_H=m_\psi=100\,\MeV$.
This bound does not depend on the DM mass, so that the constraint from $\Gamma (Z \to {\rm inv})$ is weaker than the ones from $R^{(M)}_{e/\mu}$. 

There are also one-loop corrections to $Z\to\ell\bar\ell$, 
which can be tested by the precise measurements of leptonic $Z$ boson decays. 
In our model, only the $Z$ couplings to left-handed charged leptons are modified. 
The induced deviation is so small as $|\lambda_{e,\mu}|^2 \times {\cal O}(10^{-4})$ with $m_H=200\,\GeV$. 
The deviation becomes smaller for a heavier charged scalar. 
Using the measurements of the lepton universality of $Z$ decays, 
$\Gamma(Z \to \mu^+ \mu^- )/\Gamma(Z \to e^+ e^- )=1.0009 \pm 0.0028$ and 
$\Gamma(Z \to \tau^+ \tau^- )/\Gamma(Z \to e^+ e^- )=1.0019 \pm 0.0032$ \cite{ParticleDataGroup:2024cfk}, 
we obtain the upper bounds, $|\lambda_e| < 7.42$ and $|\lambda_\mu|< 8.73$, with $m_{H_\pm}=200\,\,\GeV$, 
which are considerably weaker than the other constraints. 
Here, we required the modified ratios of the $Z\to\bar\ell\ell$ decay widths to be within $2\sigma$.

\section{Summary}
\label{sec:summary}

We have examined the role of the leptonic meson decays $M\to\ell\,\ol{\nu}_\ell$ in exploring neutrino-philic dark sectors. 
The setup we studied is based on the neutrino portal DM models, where 
the primary DM interaction to the SM sector is governed mostly by 
${\cal L}\supset \lambda_\ell \,\ol{\nu_{L}^{}}_\ell\,\phi\,\psi_R^{}$. 

We have explicitly computed the one-loop corrections to $M\to\ell\,\ol{\nu}_\ell$ 
as well as the novel three-body decay processes $M \to \ell\,\ol\psi\,\phi$ 
associated with the pertinent $M\to\ell\,\ol{\nu}_\ell$ decay process. 
The latter three-body decay influences the experimental measurements of ${\rm BR}(M\to\ell\,\ol{\nu}_\ell)$, 
as both $\psi$ and $\phi$ are invisible. 
Although the three-body decay is suppressed by the phase-space factor and the extra couplings, 
the absence of the helicity suppression in $M \to \ell\,\ol\psi\,\phi$ offers a unique opportunity to surpass the helicity suppressed $M\to\ell\,\ol{\nu}_\ell$ decay. 
Comparing the predicted lepton universality of the Kaon and pion decays 
$R^{(M)}_{e/\mu}={\rm BR}(M\to e\,\ol{\nu})/{\rm BR}(M\to\mu\,\ol{\nu})$ with the experimental measurements of that quantity, 
we have obtained the ${\cal O}(0.01)$ upper limit on the coupling $|\lambda_{e}|$ 
where the three-body decay is kinematically allowed. 
Above the kinematical thresholds, the one-loop corrections can only modify $R^{(M)}_{e/\mu}$, 
and the resulting bounds becomes weaker.
We have also shown that these strong meson decay bounds can test the thermal freeze-out scenario below ${\cal O}(100)\,\MeV$ in this model, which is otherwise unconstrained.

In the measurements of $M\to\ell\,\ol{\nu}_\ell$, 
the energy distribution of the final-state charged lepton (or equivalently the distribution of the missing mass $m_{\rm miss}$) is used to reduce background events. 
In this work, we only used the total branching ratios and did not utilize the spectral information to derive the bounds.
However, given that $M\to\ell\,\ol{\nu}_\ell$ exhibits a peak feature around $m_{\rm miss}^2=0$, 
the new physics signal from the three-body processes could be more pronounced in the high-$m_{\rm miss}^2$ tail region, offering a more powerful probe. 
To perform such a spectral analysis, the detailed information about the experimental acceptance and background distribution is imperative. 
It will be studied in a separate publication.

%%%%%%%%%%%%%%%%%%%%%%%%%%%%%%%%%%%%%%%%%%%%%%%%%%%%%%%%
\section*{Acknowledgements}
We thank Syuhei Iguro for the collaboration in the early stage of the project. 
This work is supported by the JSPS KAKENHI Grant Numbers 22K21350 and 25K17401 (S.O.) and 24K07031 (Y.O.).
The work of S.O. is also supported by an appointment to the JRG Program at the APCTP through the Science and Technology Promotion Fund and Lottery Fund of the Korean Government and by the Korean Local Governments -- Gyeongsangbuk-do Province and Pohang City.
%%%%%%%%%%%%%%%%%%%%%%%%%%%%%%%%%%%%%%%%%%%%%%%%%%%%%%%% 

\bibliographystyle{JHEP}
\bibliography{refs, WIMPs}

\providecommand{\href}[2]{#2}\begingroup\raggedright\begin{thebibliography}{10}

\bibitem{Lee:1977ua}
B.W.~Lee and S.~Weinberg, \emph{{Cosmological Lower Bound on Heavy Neutrino
  Masses}}, \href{https://doi.org/10.1103/PhysRevLett.39.165}{\emph{Phys. Rev.
  Lett.} {\bfseries 39} (1977) 165}.

\bibitem{Hut:1977zn}
P.~Hut, \emph{{Limits on Masses and Number of Neutral Weakly Interacting
  Particles}}, \href{https://doi.org/10.1016/0370-2693(77)90139-3}{\emph{Phys.
  Lett. B} {\bfseries 69} (1977) 85}.

\bibitem{Sato:1977ye}
K.~Sato and M.~Kobayashi, \emph{{Cosmological Constraints on the Mass and the
  Number of Heavy Lepton Neutrinos}},
  \href{https://doi.org/10.1143/PTP.58.1775}{\emph{Prog. Theor. Phys.}
  {\bfseries 58} (1977) 1775}.

\bibitem{Vysotsky:1977pe}
M.I.~Vysotsky, A.D.~Dolgov and Y.B.~Zeldovich, \emph{{Cosmological Restriction
  on Neutral Lepton Masses}}, {\emph{JETP Lett.} {\bfseries 26} (1977) 188}.

\bibitem{Bernstein:1985th}
J.~Bernstein, L.S.~Brown and G.~Feinberg, \emph{{The Cosmological Heavy
  Neutrino Problem Revisited}},
  \href{https://doi.org/10.1103/PhysRevD.32.3261}{\emph{Phys. Rev. D}
  {\bfseries 32} (1985) 3261}.

\bibitem{Boehm:2003hm}
C.~Boehm and P.~Fayet, \emph{{Scalar dark matter candidates}},
  \href{https://doi.org/10.1016/j.nuclphysb.2004.01.015}{\emph{Nucl. Phys. B}
  {\bfseries 683} (2004) 219}
  [\href{https://arxiv.org/abs/hep-ph/0305261}{{\ttfamily hep-ph/0305261}}].

\bibitem{Fayet:2004bw}
P.~Fayet, \emph{{Light spin 1/2 or spin 0 dark matter particles}},
  \href{https://doi.org/10.1103/PhysRevD.70.023514}{\emph{Phys. Rev. D}
  {\bfseries 70} (2004) 023514}
  [\href{https://arxiv.org/abs/hep-ph/0403226}{{\ttfamily hep-ph/0403226}}].

\bibitem{Galison:1983pa}
P.~Galison and A.~Manohar, \emph{{TWO Z's OR NOT TWO Z's?}},
  \href{https://doi.org/10.1016/0370-2693(84)91161-4}{\emph{Phys. Lett. B}
  {\bfseries 136} (1984) 279}.

\bibitem{Holdom:1985ag}
B.~Holdom, \emph{{Two U(1)'s and Epsilon Charge Shifts}},
  \href{https://doi.org/10.1016/0370-2693(86)91377-8}{\emph{Phys. Lett. B}
  {\bfseries 166} (1986) 196}.

\bibitem{Holdom:1986eq}
B.~Holdom, \emph{{Searching for $\epsilon$ Charges and a New U(1)}},
  \href{https://doi.org/10.1016/0370-2693(86)90470-3}{\emph{Phys. Lett. B}
  {\bfseries 178} (1986) 65}.

\bibitem{Foot:1991kb}
R.~Foot and X.-G.~He, \emph{{Comment on Z Z-prime mixing in extended gauge
  theories}}, \href{https://doi.org/10.1016/0370-2693(91)90901-2}{\emph{Phys.
  Lett. B} {\bfseries 267} (1991) 509}.

\bibitem{Pospelov:2007mp}
M.~Pospelov, A.~Ritz and M.B.~Voloshin, \emph{{Secluded WIMP Dark Matter}},
  \href{https://doi.org/10.1016/j.physletb.2008.02.052}{\emph{Phys. Lett. B}
  {\bfseries 662} (2008) 53} [\href{https://arxiv.org/abs/0711.4866}{{\ttfamily
  0711.4866}}].

\bibitem{Arkani-Hamed:2008hhe}
N.~Arkani-Hamed, D.P.~Finkbeiner, T.R.~Slatyer and N.~Weiner, \emph{{A Theory
  of Dark Matter}},
  \href{https://doi.org/10.1103/PhysRevD.79.015014}{\emph{Phys. Rev. D}
  {\bfseries 79} (2009) 015014}
  [\href{https://arxiv.org/abs/0810.0713}{{\ttfamily 0810.0713}}].

\bibitem{Schabinger:2005ei}
R.M.~Schabinger and J.D.~Wells, \emph{{A Minimal spontaneously broken hidden
  sector and its impact on Higgs boson physics at the large hadron collider}},
  \href{https://doi.org/10.1103/PhysRevD.72.093007}{\emph{Phys. Rev. D}
  {\bfseries 72} (2005) 093007}
  [\href{https://arxiv.org/abs/hep-ph/0509209}{{\ttfamily hep-ph/0509209}}].

\bibitem{Patt:2006fw}
B.~Patt and F.~Wilczek, \emph{{Higgs-field portal into hidden sectors}},
  \href{https://arxiv.org/abs/hep-ph/0605188}{{\ttfamily hep-ph/0605188}}.

\bibitem{Wells:2008xg}
J.D.~Wells, \emph{{How to Find a Hidden World at the Large Hadron Collider}},
  \href{https://arxiv.org/abs/0803.1243}{{\ttfamily 0803.1243}}.

\bibitem{Batell:2009yf}
B.~Batell, M.~Pospelov and A.~Ritz, \emph{{Probing a Secluded U(1) at
  B-factories}}, \href{https://doi.org/10.1103/PhysRevD.79.115008}{\emph{Phys.
  Rev. D} {\bfseries 79} (2009) 115008}
  [\href{https://arxiv.org/abs/0903.0363}{{\ttfamily 0903.0363}}].

\bibitem{Weihs:2011wp}
E.~Weihs and J.~Zurita, \emph{{Dark Higgs Models at the 7 TeV LHC}},
  \href{https://doi.org/10.1007/JHEP02(2012)041}{\emph{JHEP} {\bfseries 02}
  (2012) 041} [\href{https://arxiv.org/abs/1110.5909}{{\ttfamily 1110.5909}}].

\bibitem{Kim:2008pp}
Y.G.~Kim, K.Y.~Lee and S.~Shin, \emph{{Singlet fermionic dark matter}},
  \href{https://doi.org/10.1088/1126-6708/2008/05/100}{\emph{JHEP} {\bfseries
  05} (2008) 100} [\href{https://arxiv.org/abs/0803.2932}{{\ttfamily
  0803.2932}}].

\bibitem{Pospelov:2011yp}
M.~Pospelov and A.~Ritz, \emph{{Higgs decays to dark matter: beyond the minimal
  model}}, \href{https://doi.org/10.1103/PhysRevD.84.113001}{\emph{Phys. Rev.
  D} {\bfseries 84} (2011) 113001}
  [\href{https://arxiv.org/abs/1109.4872}{{\ttfamily 1109.4872}}].

\bibitem{Nomura:2008ru}
Y.~Nomura and J.~Thaler, \emph{{Dark Matter through the Axion Portal}},
  \href{https://doi.org/10.1103/PhysRevD.79.075008}{\emph{Phys. Rev. D}
  {\bfseries 79} (2009) 075008}
  [\href{https://arxiv.org/abs/0810.5397}{{\ttfamily 0810.5397}}].

\bibitem{Dolan:2014ska}
M.J.~Dolan, F.~Kahlhoefer, C.~McCabe and K.~Schmidt-Hoberg, \emph{{A taste of
  dark matter: Flavour constraints on pseudoscalar mediators}},
  \href{https://doi.org/10.1007/JHEP03(2015)171}{\emph{JHEP} {\bfseries 03}
  (2015) 171} [\href{https://arxiv.org/abs/1412.5174}{{\ttfamily 1412.5174}}].

\bibitem{Gola:2021abm}
S.~Gola, S.~Mandal and N.~Sinha, \emph{{ALP-portal majorana dark matter}},
  \href{https://doi.org/10.1142/S0217751X22501317}{\emph{Int. J. Mod. Phys. A}
  {\bfseries 37} (2022) 2250131}
  [\href{https://arxiv.org/abs/2106.00547}{{\ttfamily 2106.00547}}].

\bibitem{Fitzpatrick:2023xks}
P.J.~Fitzpatrick, Y.~Hochberg, E.~Kuflik, R.~Ovadia and Y.~Soreq, \emph{{Dark
  matter through the axion-gluon portal}},
  \href{https://doi.org/10.1103/PhysRevD.108.075003}{\emph{Phys. Rev. D}
  {\bfseries 108} (2023) 075003}
  [\href{https://arxiv.org/abs/2306.03128}{{\ttfamily 2306.03128}}].

\bibitem{Dror:2023fyd}
J.A.~Dror, S.~Gori and P.~Munbodh, \emph{{QCD axion-mediated dark matter}},
  \href{https://doi.org/10.1007/JHEP09(2023)128}{\emph{JHEP} {\bfseries 09}
  (2023) 128} [\href{https://arxiv.org/abs/2306.03145}{{\ttfamily
  2306.03145}}].

\bibitem{Darme:2020sjf}
L.~Darm\'e, F.~Giacchino, E.~Nardi and M.~Raggi, \emph{{Invisible decays of
  axion-like particles: constraints and prospects}},
  \href{https://doi.org/10.1007/JHEP06(2021)009}{\emph{JHEP} {\bfseries 06}
  (2021) 009} [\href{https://arxiv.org/abs/2012.07894}{{\ttfamily
  2012.07894}}].

\bibitem{Kamada:2017tsq}
A.~Kamada, H.~Kim and T.~Sekiguchi, \emph{{Axionlike particle assisted strongly
  interacting massive particle}},
  \href{https://doi.org/10.1103/PhysRevD.96.016007}{\emph{Phys. Rev. D}
  {\bfseries 96} (2017) 016007}
  [\href{https://arxiv.org/abs/1704.04505}{{\ttfamily 1704.04505}}].

\bibitem{Hochberg:2018rjs}
Y.~Hochberg, E.~Kuflik, R.~Mcgehee, H.~Murayama and K.~Schutz, \emph{{Strongly
  interacting massive particles through the axion portal}},
  \href{https://doi.org/10.1103/PhysRevD.98.115031}{\emph{Phys. Rev. D}
  {\bfseries 98} (2018) 115031}
  [\href{https://arxiv.org/abs/1806.10139}{{\ttfamily 1806.10139}}].

\bibitem{Bharucha:2022lty}
A.~Bharucha, F.~Br\"ummer, N.~Desai and S.~Mutzel, \emph{{Axion-like particles
  as mediators for dark matter: beyond freeze-out}},
  \href{https://doi.org/10.1007/JHEP02(2023)141}{\emph{JHEP} {\bfseries 02}
  (2023) 141} [\href{https://arxiv.org/abs/2209.03932}{{\ttfamily
  2209.03932}}].

\bibitem{Ghosh:2023tyz}
D.K.~Ghosh, A.~Ghoshal and S.~Jeesun, \emph{{Axion-like particle (ALP) portal
  freeze-in dark matter confronting ALP search experiments}},
  \href{https://arxiv.org/abs/2305.09188}{{\ttfamily 2305.09188}}.

\bibitem{Boehm:2013jpa}
C.~Boehm, M.J.~Dolan and C.~McCabe, \emph{{A Lower Bound on the Mass of Cold
  Thermal Dark Matter from Planck}},
  \href{https://doi.org/10.1088/1475-7516/2013/08/041}{\emph{JCAP} {\bfseries
  08} (2013) 041} [\href{https://arxiv.org/abs/1303.6270}{{\ttfamily
  1303.6270}}].

\bibitem{Batell:2017cmf}
B.~Batell, T.~Han, D.~McKeen and B.~Shams Es~Haghi, \emph{{Thermal Dark Matter
  Through the Dirac Neutrino Portal}},
  \href{https://doi.org/10.1103/PhysRevD.97.075016}{\emph{Phys. Rev. D}
  {\bfseries 97} (2018) 075016}
  [\href{https://arxiv.org/abs/1709.07001}{{\ttfamily 1709.07001}}].

\bibitem{McKeen:2018pbb}
D.~McKeen and N.~Raj, \emph{{Monochromatic dark neutrinos and boosted dark
  matter in noble liquid direct detection}},
  \href{https://doi.org/10.1103/PhysRevD.99.103003}{\emph{Phys. Rev. D}
  {\bfseries 99} (2019) 103003}
  [\href{https://arxiv.org/abs/1812.05102}{{\ttfamily 1812.05102}}].

\bibitem{Blennow:2019fhy}
M.~Blennow, E.~Fernandez-Martinez, A.~Olivares-Del~Campo, S.~Pascoli,
  S.~Rosauro-Alcaraz and A.V.~Titov, \emph{{Neutrino Portals to Dark Matter}},
  \href{https://doi.org/10.1140/epjc/s10052-019-7060-5}{\emph{Eur. Phys. J. C}
  {\bfseries 79} (2019) 555}
  [\href{https://arxiv.org/abs/1903.00006}{{\ttfamily 1903.00006}}].

\bibitem{Biswas:2021kio}
A.~Biswas, D.~Borah and D.~Nanda, \emph{{Light Dirac neutrino portal dark
  matter with observable \ensuremath{\Delta}Neff}},
  \href{https://doi.org/10.1088/1475-7516/2021/10/002}{\emph{JCAP} {\bfseries
  10} (2021) 002} [\href{https://arxiv.org/abs/2103.05648}{{\ttfamily
  2103.05648}}].

\bibitem{Li:2022bpp}
S.-P.~Li and X.-J.~Xu, \emph{{Dark matter produced from right-handed
  neutrinos}}, \href{https://doi.org/10.1088/1475-7516/2023/06/047}{\emph{JCAP}
  {\bfseries 06} (2023) 047}
  [\href{https://arxiv.org/abs/2212.09109}{{\ttfamily 2212.09109}}].

\bibitem{Okawa:2020jea}
S.~Okawa and Y.~Omura, \emph{{Light mass window of lepton portal dark matter}},
  \href{https://doi.org/10.1007/JHEP02(2021)231}{\emph{JHEP} {\bfseries 02}
  (2021) 231} [\href{https://arxiv.org/abs/2011.04788}{{\ttfamily
  2011.04788}}].

\bibitem{Yuksel:2007ac}
H.~Yuksel, S.~Horiuchi, J.F.~Beacom and S.~Ando, \emph{{Neutrino Constraints on
  the Dark Matter Total Annihilation Cross Section}},
  \href{https://doi.org/10.1103/PhysRevD.76.123506}{\emph{Phys. Rev. D}
  {\bfseries 76} (2007) 123506}
  [\href{https://arxiv.org/abs/0707.0196}{{\ttfamily 0707.0196}}].

\bibitem{PalomaresRuiz:2007eu}
S.~Palomares-Ruiz and S.~Pascoli, \emph{{Testing MeV dark matter with neutrino
  detectors}}, \href{https://doi.org/10.1103/PhysRevD.77.025025}{\emph{Phys.
  Rev. D} {\bfseries 77} (2008) 025025}
  [\href{https://arxiv.org/abs/0710.5420}{{\ttfamily 0710.5420}}].

\bibitem{Primulando:2017kxf}
R.~Primulando and P.~Uttayarat, \emph{{Dark Matter-Neutrino Interaction in
  Light of Collider and Neutrino Telescope Data}},
  \href{https://doi.org/10.1007/JHEP06(2018)026}{\emph{JHEP} {\bfseries 06}
  (2018) 026} [\href{https://arxiv.org/abs/1710.08567}{{\ttfamily
  1710.08567}}].

\bibitem{Campo:2017nwh}
A.~Olivares-Del~Campo, C.~B\oe~hm, S.~Palomares-Ruiz and S.~Pascoli,
  \emph{{Dark matter-neutrino interactions through the lens of their
  cosmological implications}},
  \href{https://doi.org/10.1103/PhysRevD.97.075039}{\emph{Phys. Rev. D}
  {\bfseries 97} (2018) 075039}
  [\href{https://arxiv.org/abs/1711.05283}{{\ttfamily 1711.05283}}].

\bibitem{Campo:2018dfh}
A.~Olivares-Del~Campo, S.~Palomares-Ruiz and S.~Pascoli, \emph{{Implications of
  a Dark Matter-Neutrino Coupling at Hyper-Kamiokande}},  in \emph{{53rd
  Rencontres de Moriond on Electroweak Interactions and Unified Theories}},
  pp.~441--444, 2018 [\href{https://arxiv.org/abs/1805.09830}{{\ttfamily
  1805.09830}}].

\bibitem{Klop:2018ltd}
N.~Klop and S.~Ando, \emph{{Constraints on MeV dark matter using neutrino
  detectors and their implication for the 21-cm results}},
  \href{https://doi.org/10.1103/PhysRevD.98.103004}{\emph{Phys. Rev. D}
  {\bfseries 98} (2018) 103004}
  [\href{https://arxiv.org/abs/1809.00671}{{\ttfamily 1809.00671}}].

\bibitem{Arguelles:2019ouk}
C.A.~Arg{\"u}elles, A.~Diaz, A.~Kheirandish, A.~Olivares-Del-Campo, I.~Safa and
  A.C.~Vincent, \emph{{Dark Matter Annihilation to Neutrinos: An Updated,
  Consistent \& Compelling Compendium of Constraints}},
  \href{https://arxiv.org/abs/1912.09486}{{\ttfamily 1912.09486}}.

\bibitem{Bell:2020rkw}
N.F.~Bell, M.J.~Dolan and S.~Robles, \emph{{Searching for Sub-GeV Dark Matter
  in the Galactic Centre using Hyper-Kamiokande}},
  \href{https://arxiv.org/abs/2005.01950}{{\ttfamily 2005.01950}}.

\bibitem{Asai:2020qlp}
K.~Asai, S.~Okawa and K.~Tsumura, \emph{{Search for $ \mathrm{U}{(1)}_{L_{\mu
  }-{L}_{\tau }} $ charged dark matter with neutrino telescope}},
  \href{https://doi.org/10.1007/JHEP03(2021)047}{\emph{JHEP} {\bfseries 03}
  (2021) 047} [\href{https://arxiv.org/abs/2011.03165}{{\ttfamily
  2011.03165}}].

\bibitem{Akita:2022lit}
K.~Akita, G.~Lambiase, M.~Niibo and M.~Yamaguchi, \emph{{Neutrino lines from
  MeV dark matter annihilation and decay in JUNO}},
  \href{https://doi.org/10.1088/1475-7516/2022/10/097}{\emph{JCAP} {\bfseries
  10} (2022) 097} [\href{https://arxiv.org/abs/2206.06755}{{\ttfamily
  2206.06755}}].

\bibitem{Restrepo:2015ura}
D.~Restrepo, A.~Rivera, M.~S{\'a}nchez-Pel{\'a}ez, O.~Zapata and W.~Tangarife,
  \emph{{Radiative Neutrino Masses in the Singlet-Doublet Fermion Dark Matter
  Model with Scalar Singlets}},
  \href{https://doi.org/10.1103/PhysRevD.92.013005}{\emph{Phys. Rev. D}
  {\bfseries 92} (2015) 013005}
  [\href{https://arxiv.org/abs/1504.07892}{{\ttfamily 1504.07892}}].

\bibitem{Esch:2016jyx}
S.~Esch, M.~Klasen, D.R.~Lamprea and C.E.~Yaguna, \emph{{Lepton flavor
  violation and scalar dark matter in a radiative model of neutrino masses}},
  \href{https://doi.org/10.1140/epjc/s10052-018-5577-7}{\emph{Eur. Phys. J. C}
  {\bfseries 78} (2018) 88} [\href{https://arxiv.org/abs/1602.05137}{{\ttfamily
  1602.05137}}].

\bibitem{Ma:2006km}
E.~Ma, \emph{{Verifiable radiative seesaw mechanism of neutrino mass and dark
  matter}}, \href{https://doi.org/10.1103/PhysRevD.73.077301}{\emph{Phys. Rev.
  D} {\bfseries 73} (2006) 077301}
  [\href{https://arxiv.org/abs/hep-ph/0601225}{{\ttfamily hep-ph/0601225}}].

\bibitem{Bonnet:2012kz}
F.~Bonnet, M.~Hirsch, T.~Ota and W.~Winter, \emph{{Systematic study of the d=5
  Weinberg operator at one-loop order}},
  \href{https://doi.org/10.1007/JHEP07(2012)153}{\emph{JHEP} {\bfseries 07}
  (2012) 153} [\href{https://arxiv.org/abs/1204.5862}{{\ttfamily 1204.5862}}].

\bibitem{Restrepo:2013aga}
D.~Restrepo, O.~Zapata and C.E.~Yaguna, \emph{{Models with radiative neutrino
  masses and viable dark matter candidates}},
  \href{https://doi.org/10.1007/JHEP11(2013)011}{\emph{JHEP} {\bfseries 11}
  (2013) 011} [\href{https://arxiv.org/abs/1308.3655}{{\ttfamily 1308.3655}}].

\bibitem{Bai:2014osa}
Y.~Bai and J.~Berger, \emph{{Lepton Portal Dark Matter}},
  \href{https://doi.org/10.1007/JHEP08(2014)153}{\emph{JHEP} {\bfseries 08}
  (2014) 153} [\href{https://arxiv.org/abs/1402.6696}{{\ttfamily 1402.6696}}].

\bibitem{Chang:2014tea}
S.~Chang, R.~Edezhath, J.~Hutchinson and M.~Luty, \emph{{Leptophilic Effective
  WIMPs}}, \href{https://doi.org/10.1103/PhysRevD.90.015011}{\emph{Phys. Rev.
  D} {\bfseries 90} (2014) 015011}
  [\href{https://arxiv.org/abs/1402.7358}{{\ttfamily 1402.7358}}].

\bibitem{Kawamura:2020qxo}
J.~Kawamura, S.~Okawa and Y.~Omura, \emph{{Current status and muon $g-2$
  explanation of lepton portal dark matter}},
  \href{https://doi.org/10.1007/JHEP08(2020)042}{\emph{JHEP} {\bfseries 08}
  (2020) 042} [\href{https://arxiv.org/abs/2002.12534}{{\ttfamily
  2002.12534}}].

\bibitem{Barbieri:2006dq}
R.~Barbieri, L.J.~Hall and V.S.~Rychkov, \emph{{Improved naturalness with a
  heavy Higgs: An Alternative road to LHC physics}},
  \href{https://doi.org/10.1103/PhysRevD.74.015007}{\emph{Phys. Rev. D}
  {\bfseries 74} (2006) 015007}
  [\href{https://arxiv.org/abs/hep-ph/0603188}{{\ttfamily hep-ph/0603188}}].

\bibitem{Pierce:2007ut}
A.~Pierce and J.~Thaler, \emph{{Natural Dark Matter from an Unnatural Higgs
  Boson and New Colored Particles at the TeV Scale}},
  \href{https://doi.org/10.1088/1126-6708/2007/08/026}{\emph{JHEP} {\bfseries
  08} (2007) 026} [\href{https://arxiv.org/abs/hep-ph/0703056}{{\ttfamily
  hep-ph/0703056}}].

\bibitem{Lundstrom:2008ai}
E.~Lundstrom, M.~Gustafsson and J.~Edsjo, \emph{{The Inert Doublet Model and
  LEP II Limits}},
  \href{https://doi.org/10.1103/PhysRevD.79.035013}{\emph{Phys. Rev. D}
  {\bfseries 79} (2009) 035013}
  [\href{https://arxiv.org/abs/0810.3924}{{\ttfamily 0810.3924}}].

\bibitem{Iguro:2022tmr}
S.~Iguro, S.~Okawa and Y.~Omura, \emph{{Light lepton portal dark matter meets
  the LHC}},  \href{https://arxiv.org/abs/2208.05487}{{\ttfamily 2208.05487}}.

\bibitem{Higuchi:2023kbt}
R.~Higuchi, S.~Iguro, S.~Okawa and Y.~Omura, \emph{{Light mass window of inert
  doublet dark matter with lepton portal interaction}},
  \href{https://doi.org/10.1103/PhysRevD.109.075007}{\emph{Phys. Rev. D}
  {\bfseries 109} (2024) 075007}
  [\href{https://arxiv.org/abs/2310.13685}{{\ttfamily 2310.13685}}].

\bibitem{Berryman:2018ogk}
J.M.~Berryman, A.~De~Gouv{\^e}a, K.J.~Kelly and Y.~Zhang,
  \emph{{Lepton-Number-Charged Scalars and Neutrino Beamstrahlung}},
  \href{https://doi.org/10.1103/PhysRevD.97.075030}{\emph{Phys. Rev. D}
  {\bfseries 97} (2018) 075030}
  [\href{https://arxiv.org/abs/1802.00009}{{\ttfamily 1802.00009}}].

\bibitem{Dev:2024ygx}
P.S.B.~Dev, D.~Kim, D.~Sathyan, K.~Sinha and Y.~Zhang, \emph{{New laboratory
  constraints on neutrinophilic mediators}},
  \href{https://doi.org/10.1016/j.physletb.2025.139765}{\emph{Phys. Lett. B}
  {\bfseries 868} (2025) 139765}
  [\href{https://arxiv.org/abs/2407.12738}{{\ttfamily 2407.12738}}].

\bibitem{Dev:2025tdv}
P.S.B.~Dev, D.~Kim, D.~Sathyan, K.~Sinha and Y.~Zhang, \emph{{New Constraints
  on Neutrino-Dark Matter Interactions: A Comprehensive Analysis}},
  \href{https://arxiv.org/abs/2507.01000}{{\ttfamily 2507.01000}}.

\bibitem{NA62:2011aa}
{\scshape NA62} collaboration, \emph{{Test of lepton flavour universality in
  $K^+ \to \ell^+ \nu$ decays}},
  \href{https://doi.org/10.1016/j.physletb.2011.02.064}{\emph{Phys. Lett. B}
  {\bfseries 698} (2011) 105}
  [\href{https://arxiv.org/abs/1101.4805}{{\ttfamily 1101.4805}}].

\bibitem{NA62:2012lny}
{\scshape NA62} collaboration, \emph{{Precision Measurement of the Ratio of the
  Charged Kaon Leptonic Decay Rates}},
  \href{https://doi.org/10.1016/j.physletb.2013.01.037}{\emph{Phys. Lett. B}
  {\bfseries 719} (2013) 326}
  [\href{https://arxiv.org/abs/1212.4012}{{\ttfamily 1212.4012}}].

\bibitem{PiENu:2015seu}
{\scshape PiENu} collaboration, \emph{{Improved Measurement of the $\pi \to
  \textrm{e} \nu$ Branching Ratio}},
  \href{https://doi.org/10.1103/PhysRevLett.115.071801}{\emph{Phys. Rev. Lett.}
  {\bfseries 115} (2015) 071801}
  [\href{https://arxiv.org/abs/1506.05845}{{\ttfamily 1506.05845}}].

\bibitem{ParticleDataGroup:2024cfk}
{\scshape Particle Data Group} collaboration, \emph{{Review of particle
  physics}}, \href{https://doi.org/10.1103/PhysRevD.110.030001}{\emph{Phys.
  Rev. D} {\bfseries 110} (2024) 030001}.

\bibitem{Finkemeier:1995gi}
M.~Finkemeier, \emph{{Radiative corrections to pi(l2) and K(l2) decays}},
  \href{https://doi.org/10.1016/0370-2693(96)01030-1}{\emph{Phys. Lett. B}
  {\bfseries 387} (1996) 391}
  [\href{https://arxiv.org/abs/hep-ph/9505434}{{\ttfamily hep-ph/9505434}}].

\bibitem{Cirigliano:2007xi}
V.~Cirigliano and I.~Rosell, \emph{{Two-loop effective theory analysis of pi
  (K) ---{¥ensuremath{>}} e anti-nu/e [gamma] branching ratios}},
  \href{https://doi.org/10.1103/PhysRevLett.99.231801}{\emph{Phys. Rev. Lett.}
  {\bfseries 99} (2007) 231801}
  [\href{https://arxiv.org/abs/0707.3439}{{\ttfamily 0707.3439}}].

\bibitem{NA62:2017ynf}
{\scshape NA62} collaboration, \emph{{Search for heavy neutrinos in $K^+
  \rightarrow \mu^+ \nu_{\mu}$ decays}},
  \href{https://doi.org/10.1016/j.physletb.2017.07.055}{\emph{Phys. Lett. B}
  {\bfseries 772} (2017) 712}
  [\href{https://arxiv.org/abs/1705.07510}{{\ttfamily 1705.07510}}].

\bibitem{NA62:2020mcv}
{\scshape NA62} collaboration, \emph{{Search for heavy neutral lepton
  production in K+ decays to positrons}},
  \href{https://doi.org/10.1016/j.physletb.2020.135599}{\emph{Phys. Lett. B}
  {\bfseries 807} (2020) 135599}
  [\href{https://arxiv.org/abs/2005.09575}{{\ttfamily 2005.09575}}].

\bibitem{NA62:2021bji}
{\scshape NA62} collaboration, \emph{{Search for $K^+$ decays to a muon and
  invisible particles}},
  \href{https://doi.org/10.1016/j.physletb.2021.136259}{\emph{Phys. Lett. B}
  {\bfseries 816} (2021) 136259}
  [\href{https://arxiv.org/abs/2101.12304}{{\ttfamily 2101.12304}}].

\bibitem{NA62:2025csa}
{\scshape NA62} collaboration, \emph{{Search for heavy neutral leptons in
  $\pi^+$ decays to positrons}},
  \href{https://arxiv.org/abs/2507.07345}{{\ttfamily 2507.07345}}.

\bibitem{PIENU:2021clt}
{\scshape PIENU} collaboration, \emph{{Search for three body pion decays
  ${\pi}^+{\to}l^+{\nu}X$}},
  \href{https://doi.org/10.1103/PhysRevD.103.052006}{\emph{Phys. Rev. D}
  {\bfseries 103} (2021) 052006}
  [\href{https://arxiv.org/abs/2101.07381}{{\ttfamily 2101.07381}}].

\end{thebibliography}\endgroup

\end{document}